\documentclass[amsmath,amssymb,a4paper,prc,final,superscriptaddress,showpacs,twocolumn]{revtex4} 

\usepackage[T1]{fontenc}
\usepackage[latin1]{inputenc}
\usepackage{bm,hyphenat,xspace}
\usepackage{epsfig,pstricks,psfig,epic,eepic,pst-coil}

\newcommand{\scl}{0.62}

\newcommand{\Eq}{Eq.}
\newcommand{\Eqs}{Eqs.}
\newcommand{\Fig}{Fig.}

\newcommand{\Ref}{Ref.}
\newcommand{\Refs}{Refs.}

\newcommand{\He}{{}^3\mathrm{He}}
\newcommand{\Hh}{{}^3\mathrm{H}}
\newcommand{\fm}{\mathrm{fm}^{-1}}
\def\Overrightarrow#1{\vbox{\ialign{##\cr
 \rightarrowfill\cr\noalign{\kern-1pt\nointerlineskip}
 $\hfil\textstyle{#1}\hfil$\cr}}}
\newcommand{\Heee}{{}^3 \vec{\mathrm{H}}\mathrm{e} (\vec{e},e')}
\newcommand {\mbf}[1]{{\mathbf{#1}}}

\newcommand {\Kpl}{\mbf{K}_{+}}

\newcommand {\Ei}{E_i \! + \!i0}
\newcommand{\mhk}{\hat{\mathbf{k}}}
\newcommand{\dEe}{dE_e(\mathbf{k}_{e_f})}

\begin{document}

\title {Three-body electrodisintegration
of the three-nucleon bound state \\ with ${\Delta}$-isobar excitation: 
Processes below pion-production threshold}

\author{A.~Deltuva} 
\thanks{On leave from Institute of Theoretical Physics and Astronomy,
Vilnius University, Vilnius 2600, Lithuania}
\email{deltuva@itp.uni-hannover.de}
\affiliation{Institut f\"ur Theoretische Physik,  Universit\"at Hannover,
  D-30167 Hannover, Germany}

\author{L.~P.~Yuan}
\affiliation{Institut f\"ur Theoretische Physik,  Universit\"at Hannover,
  D-30167 Hannover, Germany}

\author{J.~Adam~Jr.} 
\affiliation{Nuclear Physics Institute, CZ-25068 \v{R}e\v{z} near Prague, 
Czech Republic}

\author{P.~U.~Sauer}
\affiliation{Institut f\"ur Theoretische Physik,  Universit\"at Hannover,
  D-30167 Hannover, Germany}
\received{3 June 2004}


\begin{abstract}
 Electron scattering from the three-nucleon bound state with two- and
three-body disintegration is described. The description
 uses the purely nucleonic charge-dependent CD-Bonn potential and
its coupled-channel extension CD-Bonn + $\Delta$. 
 Exact solutions of three-particle  
equations are employed for the initial and final states of the reactions. 
The current has one-baryon and two-baryon contributions and couples
nucleonic with $\Delta$-isobar channels.
$\Delta$-isobar effects on the observables are isolated. 
The $\Delta$-isobar excitation yields an effective three-nucleon force 
and effective two- and three-nucleon currents beside other 
$\Delta$-isobar effects; they are mutually consistent.
\end{abstract}

 \maketitle

\section{Introduction} \label{sec:intro}

Electron scattering from the three-nucleon bound state is described
allowing for the excitation of a nucleon to a $\Delta$ isobar. 
The available energy stays below  pion-production threshold; thus,
the excitation of the $\Delta$ isobar remains virtual.
The $\Delta$ isobar is therefore considered a stable particle;
it yields an effective three-nucleon force and effective exchange currents
beside other $\Delta$-isobar effects.

The paper updates our previous
calculations~\cite{yuan:02b} of three-nucleon electron
scattering. Compared to \Ref~\cite{yuan:02b}, the description is
extended to higher energies, and three-nucleon breakup is also
included; however, energetically the description is only valid
below pion-production threshold.
Exclusive and inclusive reactions are described. 
The employed dynamics is the same as in \Ref~\cite{deltuva:04a} for photo
reactions. The underlying purely
nucleonic reference potential is CD Bonn~\cite{machleidt:01a}. Its
coupled-channel extension, called CD Bonn + $\Delta$, is employed in
this paper; it is fitted in \Ref~\cite{deltuva:03c} to the
experimental two-nucleon data up to 350~MeV nucleon lab energy; it is
as realistic as CD Bonn.
The exact solution of the three-particle scattering equations is
used for the initial- and final-state hadronic interactions. They are
solved by Chebyshev expansion of the two-baryon transition matrix as
interpolation technique \cite{deltuva:03a}; 
that technique is found highly efficient and systematic.
The employed electromagnetic (e.m.) current is structurally the same as in
\Ref~\cite{deltuva:04a} for photo reactions. It is a coupled-channel current
tuned to the used two-baryon potentials as much as possible. It contains
one- and two-baryon parts. Compared with \Ref~\cite{deltuva:04a} it is
augmented with e.m. form factors.

 An alternative description of
e.m. processes in the three-nucleon system is given in
\Refs~\cite{golak:95a,golak:95b,golak:02a}; 
\Refs~\cite{golak:95a,golak:95b,golak:02a} employ a 
different two-nucleon potential 
and a different e.m. current; nevertheless, the theoretical predictions of
\Refs~\cite{golak:95a,golak:95b,golak:02a} and of this paper
turn out to be qualitatively quite similar where comparable.

Section~\ref{sec:calc} recalls our calculational procedure and especially
stresses its improvements. Section~\ref{sec:res} presents characteristic 
results for observables; $\Delta$-isobar effects on those observables are
isolated. Section~\ref{sec:concl} gives a summary and our conclusions.

\section{\label{sec:calc} Calculational procedure}

 The kinematics of the considered processes in electron scattering 
is shown in \Fig~\ref{fig:reaction}.
The calculational procedure, including the notation, is taken over from
\Refs~\cite{deltuva:04a,yuan:02b}. 
We remind the reader shortly of that procedure in order
to point out  changes and to describe the extension to three-body electro
disintegration and to inclusive processes, not discussed in \Ref~\cite{yuan:02b}.

\subsection{Description of exclusive reactions with three-body disintegration  
\label{sec:d8s}}

\begin{figure}[!]
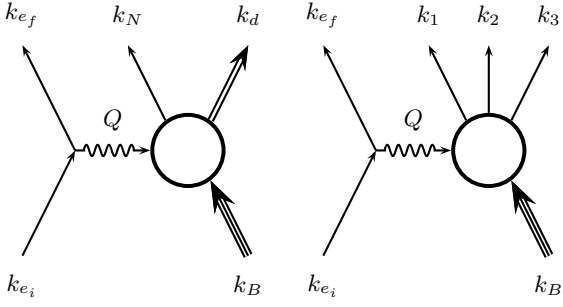

\begin{center}
\pspicture(0,0)(8.0,4.0)
\def\nucleon{\psline{->}(0,0)(-0.5,1.0)}
\def\electron{\psline{->}(-0.7,-1.4)(0,0)\psline{->}(0,0)(-0.7,1.4)}
\def\trinucl{\psline[doubleline=true,doublesep=2pt]{->}(0,0)(-0.5,1.0)
\psline(0,0)(-0.45,0.9)}
\def\deutr{\psline[doubleline=true,doublesep=1pt]{->}(0,0)(0.5,1.0)}
\def\photon{\pscoil[coilwidth=0.15cm,coilaspect=0,coilarmA=0.12cm,coilarmB=0.22cm]
{->}(0,0)(1.0,0.0)}
\multips(2.5,2)(4,0){2}{\pscircle[linewidth=1.5pt]{0.5cm}}
\multips(3.3,0.6)(4,0){2}{\trinucl} \multirput(3.3,0.2)(4,0){2}{$k_B$}
\multips(1.0,2.0)(4,0){2}{\photon}  \multirput(1.5,2.4)(4,0){2}{$Q$}
\multips(1.0,2.0)(4,0){2}{\electron} \multirput(0.3,0.2)(4,0){2}{$k_{e_i}$}
\multirput(0.3,3.8)(4,0){2}{$k_{e_f}$}
\rput(2.8,2.4){\deutr}   \rput(3.3,3.8){$k_d$}
\rput(2.2,2.4){\nucleon} \rput(1.7,3.8){$k_N$}
\multips(6.2,2.4)(3,0){1}{\nucleon} \rput(5.7,3.8){$k_1$} 
\psline{->}(6.5,2.5)(6.5,3.4)           \rput(6.5,3.8){$k_2$}
\psline{->}(6.8,2.4)(7.3,3.4)       \rput(7.3,3.8){$k_3$}
\endpspicture
 \caption{Schematic description of electrodisintegration
 of the three-nucleon bound state. Momenta are assigned to the
 particles involved. The lines for the deuteron and the three-nucleon bound
state  are drawn in a special form to indicate their compositeness.}
\label{fig:reaction}
\end{center}
\end{figure}

The $S$-matrix and the spin-averaged and spin-dependent cross sections
for two-body electrodisintegration of the trinucleon bound state are
given in \Ref~\cite{yuan:02b}.  In this subsection we add the corresponding
quantities for three-body electrodisintegration. 
The right part of
\Fig~\ref{fig:reaction} recalls the employed notation for the individual
particle momenta of the trinucleon bound state,
the three nucleons of breakup and the electron; i.e., $k_B$,
$k_j$ and $k_{e}$. They are on-mass-shell four-momenta.  The corresponding
particle energies are the zero components of those momenta, i.e.,
$k_B^0 c$, $k_j^0 c$ and $k_{e}^0 c$; they are
relativistic ones with the respective rest masses
$m_B$, $m_N$ and $m_e$, in contrast to the nonrelativistic baryonic energies
of the nonrelativistic model calculation of baryonic states
without rest masses, i.e., 
$E_B(\mbf{k}_B) = E_B + \mbf{k}_B^2/6m_N$, $E_B$ being the three-nucleon
binding energy, and $E_N(\mbf{k}_j) = \mbf{k}_j^2/2m_N$.

We give two alternative forms for the $S$-matrix elements:
\begin{subequations} \label{eq:Smat}
  \begin{align} \label{eq:Smata}
    \langle f \mbf{P}_f | S | i \mbf{P}_i \rangle  = &  
    -i(2 \pi \hbar)^{4} \delta (k_{e_f} + k_{1} + k_{2} + k_{3} - k_{e_i} - k_B)
 \nonumber \\  & \times 
    \langle s_f | M | s_i \rangle   (2 \pi \hbar)^{-9} 
 \nonumber \\  & \times 
    \big[ 2 k_{e_i}^0 c \, 2 k_B^0 c \, 2 k_{e_f}^0 c\, 
      2k_1^0 c \, 2k_2^0 c \, 2k_3^0 c \big]^{-1/2},  
    \\    \label{eq:Smatc}
    \langle f \mbf{P}_f | S | i \mbf{P}_i \rangle  = &  
    -\frac{i}{\hbar c} \, \delta \big(k_{e_f}^0 c + E_N(\mbf{k}_1)+E_N(\mbf{k}_2) 
 \nonumber \\  &
+ E_N(\mbf{k}_{3}) - k_{e_i}^0 c - E_B(\mbf{k}_B) \big)  
    \nonumber \\ & \times 
    \delta (\mbf{k}_{e_f} + \mbf{k}_{1} + \mbf{k}_{2} +
    \mbf{k}_{3} - \mbf{k}_{e_i} - \mbf{k}_B )        
    \nonumber  \\ & \times 
    \frac{1}{(2 \pi)^2}  \big[ 2 k_{e_f}^0 c \, 2 k_{e_f}^0 c \big]^{-\frac12} \,
    \nonumber  \\ & \times 
    \bar{u} (\mbf{k}_{e_f} s_{e_f}) \gamma_{\mu}  u(\mbf{k}_{e_i} s_{e_i}) \,
    \frac{4 \pi e_p^2 }{(k_{e_f} - k_{e_i})^2} 
    \nonumber \\   &   \times 
    \frac{1}{e_p c} \langle \psi^{(-)}_0 (\mbf{p}_f \mbf{q}_f) \nu_{0f} |
    j^{\mu} (\mbf{Q}, \Kpl)  | B \rangle. 
  \end{align}
\end{subequations}
Equation~\eqref{eq:Smata} introduces a covariant form, whereas 
\Eq~\eqref{eq:Smatc} is the noncovariant quantum mechanical
realization of it. 
$\mbf{P}$ is the total momentum including the one of the electron,
$(\mbf{p} \mbf{q} \mbf{K})$ the Jacobi momenta of the three baryons according to
\Ref~\cite{nemoto:98a}; $\Kpl = \mbf{K}_i +  \mbf{K}_f$;
$i$ and $f$ indicate the initial and final states of the reaction. 
$ u (\mbf{k} s)$ is the Dirac spinor of the electron 
with positive energy in the normalization  
$ \bar{u}(\mbf{k} s') u(\mbf{k} s) = m_e c^2 \delta_{s's}$;  
$ \langle s_f | M | s_i \rangle $ is the singularity-free matrix element for 
three-nucleon electrodisintegration, from which the differential cross section
 is obtained. Its dependence on the spin projections $s_{e_i} $ and  
${\mathcal M}_{B}$ of electron and trinucleon bound state in the initial channel,
 collectively described by $s_i$, and on the spin projections $s _{e_f}$ and
 $m_{s_f}$ of electron and 
nucleons  in the final channel, collectively described by $s_f$, are explicitly 
indicated. $\langle s_f | M | s_i \rangle $ is Lorentz-invariant  in a 
relativistic description and can therefore be calculated in any frame. 
However, when calculated  according to \Eq~\eqref{eq:Smatc}
in the framework of nonrelativistic quantum mechanics,
$\langle s_f | M | s_i \rangle $ loses that property of being a Lorentz scalar.

The lab cross section takes the following compact form
\begin{subequations} \label{eq:d5S}
  \begin{gather}
    \begin{split} \label{eq:d5s}
      d^8 \sigma_{i \to f} = & 
       \big | \langle s_f | M | s_i \rangle \big |^2
       \mathrm{fps} \: \dEe \, d^2 \mhk_{e_f} \, 
       d S \, d^2 \mhk_{1} \, d^2 \mhk_{2}
    \end{split}
\end{gather}
with the abbreviation $\mathrm{fps}$  for a phase-space factor; 
in the lab frame $\mathrm{fps}$ is
  \begin{align}  \label{eq:fpsr}
    \mathrm{fps} = & \frac{k_{e_f}^0} {(2 \pi \hbar)^{8}
	64 c^7 k_{e_i}^0  m_B }\, \mbf{k}_1^2  \mbf{k}_2^2  
    \nonumber \\  & \times 
    \big \{ \mbf{k}_1^2   \big[ |\mbf{k}_2| (k_2^0 + k_3^0) 
      - k_2^0 \mhk_2 \cdot (\mbf{Q} \! - \! \mbf{k}_1) \big]^2  
    \nonumber \\  & 
    + \mbf{k}_2^2 \big[ |\mbf{k}_1| (k_1^0+k_3^0)  - 
      k_1^0 \mhk_1 \cdot (\mbf{Q} \! - \! \mbf{k}_2) \big]^2 \big \}^{-\frac12}, \\
 \label{eq:fpsn} 
    \mathrm{fps} = & \frac{k_{e_f}^0}{(2 \pi \hbar)^{8} 64 c^8 k_{e_i}^0 m_N m_B }
    \, \mbf{k}_1^2  \mbf{k}_2^2   
    \nonumber \\  & \times 
    \big \{ \mbf{k}_1^2 \big[ 2|\mbf{k}_2|  - 
      \mhk_2 \cdot (\mbf{Q}  \! - \! \mbf{k}_1) \big]^2  
    \nonumber \\  & 
     + \mbf{k}_2^2 \big[ 2|\mbf{k}_1|  - 
      \mhk_1 \cdot (\mbf{Q}  \! - \! \mbf{k}_2) \big]^2 \big \}^{-\frac12}. 
  \end{align}
\end{subequations}
Equation~\eqref{eq:fpsn} is the nonrelativistic version of \Eq~\eqref{eq:fpsr};
$dS$ is the element of arclength $S$ as used in \Ref~\cite{deltuva:04a}.
  The cross section~\eqref{eq:d5s} is still spin-dependent.
  The spin-averaged eightfold differential cross section is
\begin{gather} \label{eq:d5s-av}
  \frac{\overline{d^8 \sigma}}{ \dEe  d^2  \mhk_{e_f}\, dS
     \, d^2 \mhk_{1} \, d^2 \mhk_{2}} = \nonumber \\
  \frac 14 \sum_{s_f s_i}   \frac {d^8 \sigma_{i\to f}} 
	{\dEe d^2 \mhk_{e_f} d S \, d^2 \mhk_{1} \, d^2 \mhk_{2} } ;
\end{gather}
in  figures it is denoted by 
$d^8 \sigma / dE_e d\Omega_e dS d\Omega_1 d\Omega_2$, the traditional notation. 
The experimental setup determines the isospin character of 
the two detected nucleons 1 and 2; their isospin character is not followed up
in our notation.

We calculate the matrix element $\langle s_f | M | s_i \rangle$
in the lab frame using the following computational strategy.  
The strategy is in the spirit of  \Ref~\cite{deltuva:04a};
it is nonunique, since the model calculations, due to the limitations
of the underlying dynamics, miss the trinucleon binding
energy; the necessary correction for that miss has arbitrary features:

1. The experimental four-momentum transfer 
  \begin{subequations}
    \begin{align}
      \label{eq:Qe}
      Q = {} & k_{e_i} - k_{e_f} , \\
      \label{eq:Qh}
      Q = {} & k_1 + k_2 + k_3 - k_B
    \end{align}
  \end{subequations}
  determines the total energy and the total momentum of the hadronic
  part of the system in the final channel. This step is
  done using relativistic kinematics and the true experimental
  trinucleon binding energy. The experimental momentum $Q$ in the lab
  frame with $\mbf{K}_i = \mbf{k}_B=0$ determines the total momentum $\mbf{K}_f$
  and the energy $E_0 (\mbf{p}_f \mbf{q}_f \mbf{K}_f)$ of the final
  three-nucleon system in the lab frame, i.e., $\mbf{K}_f =  \mbf{Q} $ 
  and $E_0 (\mbf{p}_f \mbf{q}_f \mbf{K}_f) = E_B + Q_0 c$.
  The resulting energy $E_0 (\mbf{p}_f \mbf{q}_f \mbf{K}_f)$ of the
  final state is the true experimental one. Thus, the experimental two- and 
  three-body breakup thresholds are exactly reproduced. 

2. The matrix element $\langle s_f | M | s_i \rangle $ is
  calculated in the lab system as \emph{on-energy-shell element} under
  nonrelativistic model assumptions for hadron dynamics.
  Taking the computed trinucleon model binding
  energy $E_B$ and the average nucleon mass $m_N $, i.e., $m_N c^2 =
  938.919$~MeV, the energy transfer $Q_0$ to be used for the current  matrix 
  element results, i.e., $Q_0 c = E_0 (\mbf{p}_f \mbf{q}_f \mbf{K}_f) - E_B$;
  the three-momentum transfer $\mbf{Q}$ to be used for the 
  current  matrix element is $\mbf{Q} = \sqrt{Q_0^2+Q^2}\hat{\mbf{K}}_f$ with the
  true experimental value of $Q^2$; note that we define the square of the
  space-like  four-momentum transfer positive as $Q^2 = \mbf{Q}^2 - Q_0^2$.
  Since the model trinucleon binding energy is not the experimental one, the
  components of the resulting four-momentum transfer $Q$ do not match precisely
  their experimental values when calculating the part
  $\langle \psi^{(-)}_0 (\mbf{p}_f \mbf{q}_f) \nu_{0f} |
  j^{\mu} (\mbf{Q}_{}, \Kpl ) | B \rangle$ of the matrix element
  $\langle s_f | M |s_i \rangle$ according to \Eq~\eqref{eq:Mampl} below;
  this strategy is chosen in order to preserve the experimental
  $Q^2$ as in photo reactions \cite{deltuva:04a}.
  The internal three-nucleon energy part of the final state is
  ${\mbf{p}_f^2}/{m_N} + {3\mbf{q}_f^2}/{4 m_N} = 
  E_0 (\mbf{p}_f  \mbf{q}_f \mbf{K}_f) - {\mbf{K}_f^2}/{6 m_N}$.

3. The lab cross section is calculated nonrelativistically; it is
constructed from the following form of the matrix element
\begin{gather} \label{eq:Mampl}
  \begin{align} 
    \langle s_f | M | s_i \rangle = & \frac{\hbar}{c} (2 \pi \hbar)^{3} 
    \bar{u} (\mbf{k}_{e_f} s_{e_f}) \gamma_{\mu} u (\mbf{k}_{e_i} s_{e_i})
    \frac{4 \pi e_p^2 }{(k_{e_f} - k_{e_i})^2 } 
    \nonumber \\   &   \times  \frac{1}{e_p c}    
    \langle \psi^{(-)}_0 (\mbf{p}_f \mbf{q}_f) \nu_{0f} |
    j^{\mu} (\mbf{Q}, \mbf{Q}) | B \rangle
 \nonumber \\   &   \times  [2m_N  c^2 ]^{3/2} [ 2m_B c^2 ]^{1/2} 
  \end{align}
\end{gather}
and from the nonrelativistic phase-space factor $\mathrm{fps}$ in the 
form~\eqref{eq:fpsn}.
As discussed in \Ref~\cite{deltuva:04a} one could choose the hadronic 
kinematics nonrelativistically
for the dynamic matrix element $\langle s_f | M | s_i \rangle $ on one side
and relativistically for the kinematical factors on the other side. 
That split calculational strategy 
can be carried out with ease for the observables of exclusive
processes.  However, when total cross sections 
or inelastic structure functions in inclusive processes are calculated,
we resort to a particular technical scheme as already 
described in  \Ref~\cite{deltuva:04a} for the total photo cross section:
The energy-conserving $\delta$-function in the phase-space 
element is rewritten as imaginary part of 
the full resolvent and that full resolvent has to be made consistent 
with the employed nonrelativistic dynamics of the model calculations.
Thus, the split calculational strategy, developed in \Ref~\cite{yuan:02b},
cannot be carried through for total cross sections and inelastic
structure functions.
We therefore do not use it in our \emph{standard calculational procedure};
we use it only for exclusive cross sections
when testing the validity of the employed nonrelativistic kinematics.

\subsection{Description of inclusive reactions \label{sec:RF}}

We assume that the electron beam is polarized with the electron
helicity $h_e$ and that the trinucleon target is polarized according to the
polarization vector $\mbf{n}_B = (\sin \theta_B \cos \varphi_B,
\sin \theta_B \sin \varphi_B, \cos \theta_B)$; the angles are taken with
respect to the direction $\hat{\mbf{Q}}$. 
We use the same definition of coordinate axes as \Ref~\cite{yuan:02b}.

The inclusive spin-dependent differential cross section has the form
\begin{gather}
  \begin{align}
    \frac{d^3 \sigma(h_e, \mbf{n}_B)}{\dEe d^2 \mhk_{e_f}}
    = {} & \sigma_{\mathrm{Mott}}
    \{v_L(Q \theta_e) R_L(Q) + v_T(Q \theta_e) R_T(Q) 
    \nonumber \\ & 
    + h_e [ v_{T'}(Q \theta_e) R_{T'}(Q) n_{Bz} 
      \nonumber \\ & + 
      v_{TL'}(Q \theta_e) R_{TL'}(Q) n_{Bx}] \}
  \end{align}
\end{gather}
with the Mott cross section $\sigma_{\mathrm{Mott}}$ and the
kinematical functions $v_{L}(Q \theta_e)$, $v_{T}(Q \theta_e)$,
$v_{T'}(Q \theta_e)$ and $v_{TL'}(Q \theta_e)$ given in \Ref~\cite{yuan:02b},
and with the inclusive response functions $R_{L}(Q)$, $R_{T}(Q)$, $R_{T'}(Q)$
and $R_{TL'}(Q)$ given in Appendix~\ref{app:RF}.
The longitudinal and transverse response functions $R_{L}(Q)$ and $R_{T}(Q)$ 
refer to a spin averaged target, $R_{T'}(Q)$ and $R_{TL'}(Q)$ are 
characteristic for the spin structure of the target.
Experiments usually measure the asymmetry $A(\mbf{n}_B)$, i.e.,
\begin{subequations}
  \begin{align}
    A (\mbf{n}_B)= {} & 
    \left[ \frac{d^3 \sigma(1, \mbf{n}_B)}{\dEe d^2 \mhk_{e_f}}
      - \frac{d^3 \sigma(-1, \mbf{n}_B)}{\dEe d^2 \mhk_{e_f}} \right]
    \Bigg/ \nonumber \\ &
    \left[ \frac{d^3 \sigma(1, \mbf{n}_B)}{\dEe d^2 \mhk_{e_f}}
      + \frac{d^3 \sigma(-1, \mbf{n}_B)}{\dEe d^2 \mhk_{e_f}} \right],
    \\
    A (\mbf{n}_B) = {} &  \frac{ v_{T'}(Q \theta_e) R_{T'}(Q) n_{Bz} + 
      v_{TL'}(Q \theta_e) R_{TL'}(Q) n_{Bx} }
    {v_L(Q \theta_e) R_L(Q) + v_T(Q \theta_e) R_T(Q) }.
\end{align}
\end{subequations}
When orienting the target spin parallel to the momentum transfer $\mbf{Q}$,
i.e., $\mbf{n}_{BT'}=(0,0,1)$, the transverse asymmetry 
$A_{T'} = A(\mbf{n}_{BT'})$ is selected;
when orienting the target spin perpendicular to the momentum transfer 
$\mbf{Q}$, but in the electron scattering plane, i.e., $\mbf{n}_{BTL'}=(1,0,0)$,
the transverse-longitudinal asymmetry $A_{TL'}= A(\mbf{n}_{BTL'})$ is selected.

\section{Results \label{sec:res}}

We present results for spin-averaged and spin-dependent observables 
in electro disintegration of the three-nucleon bound state.
The presented exclusive results refer to three-body disintegration. 
Results of exclusive two-body disintegration are not shown; results for them
are given in \Ref~\cite{yuan:02b}; control calculations 
indicate that the results of \Ref~\cite{yuan:02b} do not get any essential
physics change, though the hadronic interaction and the e.m. current are 
improved compared with \Ref~\cite{yuan:02b}.

The results of this paper are based on calculations derived from the purely 
nucleonic CD-Bonn potential~\cite{machleidt:01a} and its coupled-channel 
extension~\cite{deltuva:03c}, which allows for single $\Delta$-isobar
excitation in isospin-triplet partial waves. 
The $\Delta$ isobar is considered to be a stable particle of spin and 
isospin $\frac32$ with a rest mass $m_{\Delta}c^2$ of 1232~MeV.
In contrast to the coupled-channel potential constructed previously by the 
subtraction technique~\cite{hajduk:83a} and used in the calculations of
\Ref~\cite{yuan:02b}, the new one of \Ref~\cite{deltuva:03c} 
and used in this paper is fitted properly 
to data and accounts for two-nucleon scattering data with the same quality as the
original CD-Bonn potential. We describe first the \emph{standard calculational 
procedure} which this paper follows.

The baryonic potential is taken into account in purely nucleonic and in 
nucleon-$\Delta$ partial waves up to the total two-baryon angular momentum $I=3$.
The calculations omit the Coulomb potential between charged baryons.
Nevertheless, the theoretical description is charge dependent. For reactions on
${}^3\mathrm{He}$ the proton-proton $(pp)$ and neutron-proton $(np)$ 
parts of the potentials are used, 
for reactions on ${}^3\mathrm{H}$ the neutron-neutron $(nn)$ and $np$ parts.
Assuming charge independence, the three-nucleon bound state and
nucleon-deuteron scattering states are pure states with total isospin
$\mathcal{T} = \frac12$; the three-nucleon scattering states have total
isospin $\mathcal{T} = \frac12$ and $\mathcal{T} = \frac32$, but those parts 
are not coupled by hadron dynamics. In contrast, allowing for charge dependence,
all three-baryon states have $\mathcal{T} = \frac12$ and $\mathcal{T} = \frac32$
components which are dynamically coupled. 
For hadronic reactions that coupling is found to be quantitatively important 
in the ${}^1S_0$ partial wave~\cite{deltuva:03b}; in other partial waves
the approximative treatment of charge dependence as described in 
\Ref~\cite{deltuva:03b} is found to be sufficient; it does not  couple
total isospin $\mathcal{T} = \frac12$ and $\frac32$ channels dynamically.
The same holds for the hadronic dynamics in
e.m. reactions considered in this paper:
The effect of charge dependence is dominated by the ${}^1S_0$ partial wave; 
it is seen in some particular kinematic situations, but we refrain
from discussing them in detail in this paper.
However, the calculations of e.m. reactions require  total isospin
$\mathcal{T} = \frac32$ components of scattering states
in \emph{all} considered isospin-triplet
two-baryon partial waves, since the e.m. current couples the 
$\mathcal{T} = \frac12$ and $\mathcal{T} = \frac32$ components strongly.

The  three-particle equations for the trinucleon bound state $|B \rangle$ 
and for the scattering states are solved as in \Ref~\cite{deltuva:03a};
in fact, the scattering states are calculated only implicitly as described
in Appendix~\ref{app:RF}. The resulting binding energies
of ${}^3\mathrm{He}$ are -7.941 and -8.225~MeV for CD Bonn and CD Bonn + $\Delta$,
respectively.  If the Coulomb interaction were taken into account, 
as proper for ${}^3\mathrm{He}$, the binding energies shift to
 -7.261 and -7.544~MeV, whereas the experimental value is -7.718~MeV.
  Nevertheless, we use the purely hadronic energy values
and bound-state wave functions for consistency when calculating the current 
matrix elements, since we are unable to include the Coulomb interaction 
in the scattering states.

Whereas the baryonic potential is considered up to $I=3$, the e.m.
 current is allowed to act between partial waves up to $I=6$,
the higher partial waves being created by the geometry of antisymmetrization.
The e.m. current is taken over from \Ref~\cite{deltuva:04a} augmented by
e.m. form factors. Whereas the employed current operators depend
on the three-momentum transfer $\mbf{Q}$ only, the added e.m. form factors
depend on the four-momentum transfer $Q^2 = \mbf{Q}^2 - Q_0^2$ as discussed
in Appendix A of  \Ref~\cite{deltuva:04a}, $Q_0$ being taken as the energy
transfer to the nuclear system.
The current is expanded in multipoles as described
in \Refs~\cite{oelsner:phd,deltuva:phd}; current conservation is imposed 
explicitly  by replacing the longitudinal current part by its charge part.
The technique for calculating multipole
 matrix elements is developed in \Ref~\cite{oelsner:phd}; a special
 stability problem~\cite{yuan:02a} arising in the calculation requires some 
modifications of that technique as described in \Ref~\cite{deltuva:phd}.
The electric and magnetic multipoles are calculated 
 from the one- and two-baryon parts of the spatial current;
the Siegert form of electric multipoles is \emph{not} used.
The Coulomb multipoles are calculated from diagonal single-nucleon and 
single-$\Delta$ parts of the  charge density;
 the nucleon-$\Delta$ transition contribution as well as 
two-baryon contributions are of relativistic order and are therefore omitted 
in the charge-density operator when calculating Coulomb multipoles. 

The number of considered current multipoles is limited by the maximal total 
three-baryon angular momentum $\mathcal{J}_{\mathrm{max}} = \frac{25}{2}$,
taken into account for the hadronic scattering states.
The results for the considered e.m. reactions 
 appear fully converged with respect to higher 
two-baryon angular momenta $I$, with respect to $\Delta$-isobar coupling and
with respect to higher three-baryon angular momenta $\mathcal{J}$
on the scale of accuracy which present-day experimental data require,
the exception being only exclusive observables in the vicinity of the quasielastic
peak which show poorer convergence with respect to $\mathcal{J}$.

\subsection{E.m. form factors of the three-nucleon bound state and
detailed choice of current}

The trinucleon form factors refer to elastic electron scattering.
The form factors are calculated in order to check how realistic the 
underlying current operators are for the momentum transfers required later on in 
inelastic electron scattering; they are calculated in the Breit frame,
i.e., as functions of $Q = \sqrt{Q^2} = |\mbf{Q}|$; we will make sure in the text 
that the magnitude $Q$ will not be confused with the four vector $Q$. 
As customary we give $Q$ in this subsection in units of $\fm$ with 
$1\;\fm \approx 200\;\mathrm{MeV}/c$ in contrast
to the remainder of the paper.
 The operator forms are defined in Appendix A of \Ref~\cite{deltuva:04a} 
with the hadronic parameters of the CD Bonn and CD Bonn + $\Delta$ potentials 
and with the following additional choices for
the baryonic and mesonic e.m. form factors.

We employ the recent parametrization of the nucleonic e.m. form factors
as given in \Ref~\cite{hammer:04a}; it is tuned to new form factor data
for the proton and the neutron and is therefore rather different at momentum
transfers larger than $3\,\fm$ compared with older parametrizations as those 
of \Ref~\cite{gari:86a}, used by us previously in \Refs~\cite{yuan:02b,yuan:02a}.
We take the Sachs form factors $g_E(Q^2)$ and $g_M(Q^2)$ of 
\Ref~\cite{hammer:04a} as the form factors $e(Q^2)$ and $\mu(Q^2)$ 
in Appendix A of \Ref~\cite{deltuva:04a};
in the context of the two-baryon potentials CD Bonn and CD Bonn + $\Delta$
of this paper the two-baryon exchange currents of \Eqs~(A5) -- (A7) in 
 \Ref~\cite{deltuva:04a} are used with the isovector form factors
$e^V(Q^2) = g_E^V(Q^2)$ instead of the Dirac form factor $f_1^V(Q^2)$,
used previously \cite{yuan:02b,yuan:02a} in the context of the potentials
Paris and Paris + $\Delta$. However, the two-baryon exchange currents, 
corresponding to nondiagonal meson exchanges according to \Eqs~(A5) and (A6)
of \Ref~\cite{deltuva:04a}, are used with form factors 
$f_{\rho \pi \gamma}(Q^2) = g_{\rho \pi \gamma} f_1^S(Q^2)$ and 
$f_{\omega \pi \gamma}(Q^2) = g_{\omega \pi \gamma} f_1^V(Q^2)$.
In contrast to  \Ref~\cite{deltuva:04a}, we choose for the nucleon-$\Delta$
transition form factor $g_{\Delta N}^{M1}(Q^2)$ the coupling strength as 
$g_{\Delta N}^{M1}(0) = 4.59\,\mu_N$, $\mu_N$ being the nuclear magneton.
The coupling strength is in accordance with the relation
$g_{\Delta N}^{M1}(0) = \frac32 G_M^\ast(0)$ to the transition 
magnetic moment $ G_M^\ast(0)$ and with its experimental value
$ G_M^\ast(0) = 3.06\,\mu_N$ of \Ref~\cite{kamalov:01a},
the experimental value being a bit larger than the quark model value
$2.63\,\mu_N$. The momentum-transfer dependence of $g_{\Delta N}^{M1}(Q^2)$
is taken as
$g_{\Delta N}^{M1}(Q^2)/g_{\Delta N}^{M1}(0)= 
e^{-\gamma Q^2}/(1+Q^2/\Lambda_{\Delta N}^2)^{2}$
with $\gamma=0.21\,(\mathrm{GeV/c})^{-2}$ and 
$\Lambda_{\Delta N}^2 = 0.71\,(\mathrm{GeV}/c)^2$
according to \Fig~2 of \Ref~\cite{kamalov:01a};
the momentum-dependent fall of the form factor $g_{\Delta N}^{M1}(Q^2)$ is
slightly faster than for a dipole.

\begin{figure}[!]
\includegraphics[scale=\scl]{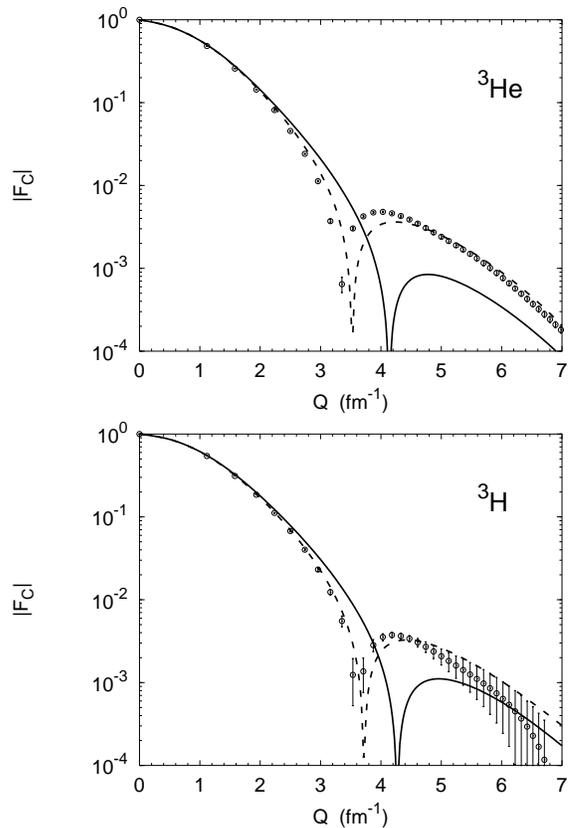}
\caption{\label{fig:fc}
Charge form factors $F_C$ of $\He$ and $\Hh$ as function of
momentum transfer $Q$.
Results of the coupled-channel potential with $\Delta$-isobar excitation 
without (solid curves) and with selected relativistic charge operator
corrections (dashed curves) are compared.
The results of the purely nucleonic CD-Bonn potential are almost
indistinguishable from the respective results of CD Bonn + $\Delta$.
The experimental data are from \Ref~\cite{sick:01a}.}
\end{figure}

Figure~\ref{fig:fc} shows the trinucleon charge form factors.
The relativistic operator corrections as given in \Eqs~(A11) of 
\Ref~\cite{deltuva:04a} and the additional corrections of $\rho$ and $\omega$
exchange of \Ref~\cite{carlson:98a} are necessary to account for the data 
at least roughly at larger momentum transfers. 
Those relativistic corrections are, however, 
omitted in our \emph{standard calculational procedure} for electrodisintegration.
Most disintegration processes, considered in this paper, require the current 
up to momentum transfers $|\mbf{Q} | \approx 2.5\,\fm$; 
in that kinematic regime the employed relativistic corrections are still small. 
However, even in that limited kinematic regime the predictions based on
that \emph{standard calculational procedure} show more deviations from data
with increasing momentum transfer. The found agreement between data and the 
theoretical predictions with relativistic corrections
for the trinucleon charge form factors is comparable with the results of 
\Refs~\cite{henning:95a,marcucci:98a}, based on other baryonic potentials.
In contrast to the relativistic corrections, the nonrelativistic
$\Delta$-isobar effect on the charge form factors is minute and therefore
not separately shown in \Fig~\ref{fig:fc}.

\begin{figure}[!]
\includegraphics[scale=\scl]{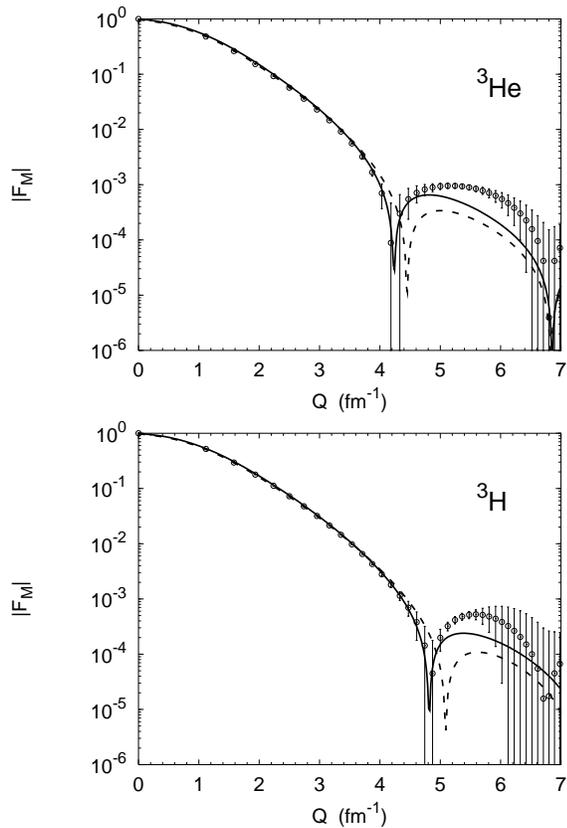}
\caption{\label{fig:fm}
Magnetic form factors $F_M$ of $\He$ and $\Hh$ as function of
momentum transfer $Q$.
Results of the coupled-channel potential with $\Delta$-isobar excitation 
(solid curves) are compared with reference results of the purely nucleonic 
CD-Bonn potential (dashed curves).
The experimental data are from \Ref~\cite{sick:01a}.
Both, data and theoretical predictions, are divided by the experimental
magnetic moments as given in Table~\ref{tab:mm}.}
\end{figure}

The trinucleon magnetic moments are given in Table~\ref{tab:mm} and
the  magnetic form factors are shown in \Fig~\ref{fig:fm}.
The agreement between data and theoretical predictions is quite satisfactory
for the magnetic moments and for the form factors up to $Q=5\,\fm$;
beyond $Q=5\,\fm$ the predicted form factors are too small in magnitude and
have a shape not consistent with the data around the secondary maximum.
In contrast to the charge, exchange 
corrections of the spatial current are of nonrelativistic order and
they contribute already at momentum transfers relevant for the considered
disintegration processes; they are therefore fully included in our
\emph{standard calculational procedure}.
In the context of the potentials CD Bonn and CD Bonn + $\Delta$, 
the use of the isovector form factors $g_E^V(Q^2)$ for diagonal 
meson-exchange currents is absolutely necessary; the use of $f_1^V(Q^2)$ 
instead moves the first minima out, i.e., beyond $7\,\fm$.

At the larger momentum transfers $Q > 3\,\fm$ 
we note a sensitivity of the theoretical predictions
for the charge and magnetic form factors upon the
parametrization of the underlying charge and current operators,
especially on the e.m. form factors of baryons.
Furthermore, as already discussed in \Ref~\cite{deltuva:04a},
the match between hadronic and e.m. dynamics has deficiencies,
i.e., the two-baryon potentials being nonlocal, whereas the employed e.m. currents
being local; also the use of a nonrelativistic description 
of the hadron dynamics at those momentum transfers is questionable.
However, the observed theoretical uncertainties and the discrepancies with data,
occurring at larger momentum transfers, are not relevant for most
disintegration processes, considered in this paper.

\begin{table}[htbp]
  \centering
\begin{ruledtabular}
  \begin{tabular}{{l}*{2}{c}}
 & $\mu(\He)$ & $\mu(\Hh)$ \\ \hline
CD Bonn & $-2.073$ & 2.906    \\
CD Bonn + $\Delta$ & $-2.139$ & 2.970 \\
Experiment & $-2.127$ & 2.979  
  \end{tabular}
\end{ruledtabular}
  \caption{ \label{tab:mm}
Magnetic moments $\mu$ of $\He$ and $\Hh$ in units of the nuclear 
magneton $\mu_N$. }
\end{table}

\subsection{Exclusive three-nucleon breakup \label{sec:3Nexcl}}

\begin{figure}[!]
\begin{center}
\includegraphics[scale=\scl]{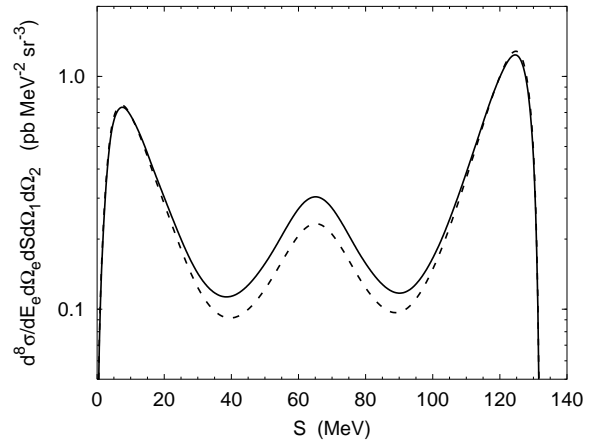}
\end{center}
\caption{\label{fig:e3HR}
Eightfold differential cross section of three-body electrodisintegration 
of $\He$, i.e., $\He(e,e'pp)n$,
at 390~MeV electron lab energy as a function of the arclength $S$ along the 
kinematical curve. The electron
scattering angle, the momentum and energy transfer are
 $\theta_e = 39.7^{\circ}$, $\;|\mbf{Q}| = 250.2\;\mathrm{MeV}/c$ and 
$\;Q_0 = 113\;\mathrm{MeV}/c$, respectively.
The observable  refers to the configuration
$(30^{\circ},180^{\circ},45^{\circ},180^{\circ})$;
the angles are given with respect to the direction of the incoming electron;
the notation is standard, e.g., explained in \Ref~\cite{deltuva:phd}.
Results of the coupled-channel potential with $\Delta$-isobar excitation 
(solid curves) are compared with reference results of the purely nucleonic 
CD-Bonn potential (dashed curves).}
\end{figure}

To the best of our knowledge, there are no fully exclusive experimental data 
of three-nucleon breakup in the considered energy regime. 
As in hadronic and photo reactions we observe more significant
$\Delta$-isobar effects at higher energies.
Figure~\ref{fig:e3HR} presents sample results for
the spin-averaged eightfold differential cross section of the three-body electro
disintegration of $\He$ 
with a moderate $\Delta$-isobar effect.

\begin{figure}[!]
\begin{center}
\includegraphics[scale=\scl]{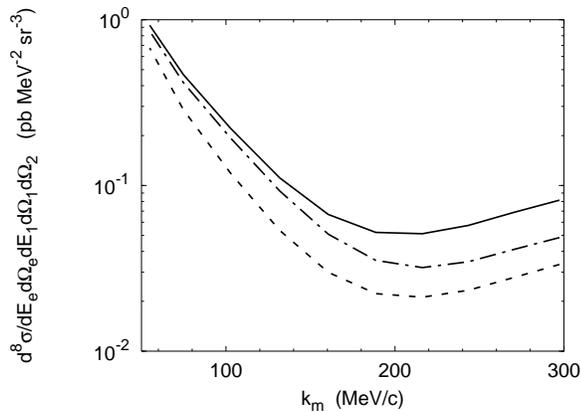}
\end{center}
\caption{\label{fig:LQ}
Eightfold differential cross section of $\He(e,e'pp)n$ reaction
at 563.7~MeV electron lab energy as a function of the magnitude of
missing momentum $\mbf{k}_m = \mbf{Q} - \mbf{k}_1 - \mbf{k}_2 $.
The electron scattering angle, the momentum and energy transfer are
 $\theta_e = -27.72^{\circ}$, $\;|\mbf{Q}| = 305\;\mathrm{MeV}/c$ and 
$\;Q_0 = 220\;\mathrm{MeV}/c$, respectively.
The observable  refers to the configuration
$(53.8^{\circ},0.0^{\circ},92.9^{\circ},180.0^{\circ})$.
Results of the standard coupled-channel potential with $\Delta$-isobar excitation 
(solid curve) and of its modified version (dash-dotted curve)
are compared with reference results of the purely nucleonic 
CD-Bonn potential (dashed curve).}
\end{figure}

Reference~\cite{groep:00a} presents results for the eightfold differential 
cross section 
${d^8 \sigma}/{ d E_e d\Omega_e \, dE_1 \, d\Omega_1 \, d\Omega_2}$
averaged over a rather large experimental detection volume. However, the 
excitation energy in the experiment of \Ref~\cite{groep:00a} is well
above pion-production threshold. Thus, the theoretical predictions of any
model neglecting pionic channels as our potential CD Bonn + $\Delta$
should be taken with severe caution. Nevertheless,
we present our results for that higher energy in \Fig~\ref{fig:LQ};
we use the representation of \Ref~\cite{groep:00a}, but for simplicity we do not
perform the averaging. We note large $\Delta$-isobar effects in particular
kinematical regimes where the purely nucleonic  calculations presented in \Fig~7 
of \Ref~\cite{groep:00a} clearly underestimate the data; the inclusion
of the $\Delta$ isobar may therefore be able to reduce that discrepancy.
However, we emphasize that the employed potentials CD Bonn and CD Bonn + $\Delta$
are unrealistic above pion-production threshold; in contrast to CD Bonn,
the coupled-channel potential CD Bonn + $\Delta$ yields inelasticities, 
but they show clearly unphysical, resonating and therefore unwanted structures
in the ${}^1D_2$ two-nucleon partial wave as already demonstrated in 
\Ref~\cite{deltuva:03c}, casting serious doubts on the size of the calculated
$\Delta$-isobar effect in \Fig~\ref{fig:LQ}.
Thus, a modified version of CD Bonn + $\Delta$ with more realistic phase shifts
above pion-production threshold is developed for exploratory reasons 
and also used for the predictions in \Fig~\ref{fig:LQ}; it yields an expected 
reduction of the $\Delta$-isobar effect, though the effect remains rather 
strong. The modified version of CD Bonn + $\Delta$ works with a reduced coupling 
strength $g_{\sigma \Delta \Delta}$ of the $\sigma$ meson to the $\Delta$ isobar.
The quality of the fit to the two-nucleon scattering data up to 350~MeV
nucleon lab energy remains practically unchanged, but, unfortunately, 
the beneficial $\Delta$-isobar effect on trinucleon binding gets almost 
completely lost --  the resulting binding energy of $\He$ including the
Coulomb interaction is -7.329~MeV.
Other modification schemes of CD Bonn + $\Delta$ have not been tried yet.
Anyhow, the data of \Ref~\cite{groep:00a} deserve a theoretical description
with a two-baryon potential, extended realistically above pion-production 
threshold.

\subsection{Inclusive response functions}

\begin{figure}[!]
\begin{center}
\includegraphics[scale=\scl]{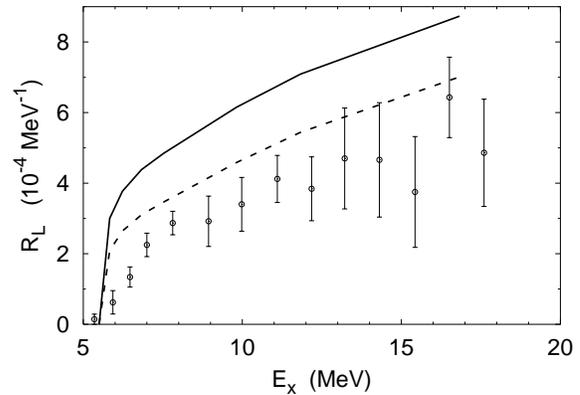}
\end{center}
\caption{\label{fig:RLt}
$\He$ inclusive longitudinal response $R_L$ near threshold
for the momentum transfer $|\mbf{Q}| = 487\;\mathrm{MeV}/c$ 
as function of the excitation energy  $E_x$. 
Results of the coupled-channel potential with $\Delta$-isobar excitation 
without (solid curves) and with selected relativistic charge operator
corrections (dashed curves) are compared.
The experimental data are from \Ref~\cite{retzlaff:94a}.}
\end{figure}

\begin{figure}[!]
\begin{center}
\includegraphics[scale=\scl]{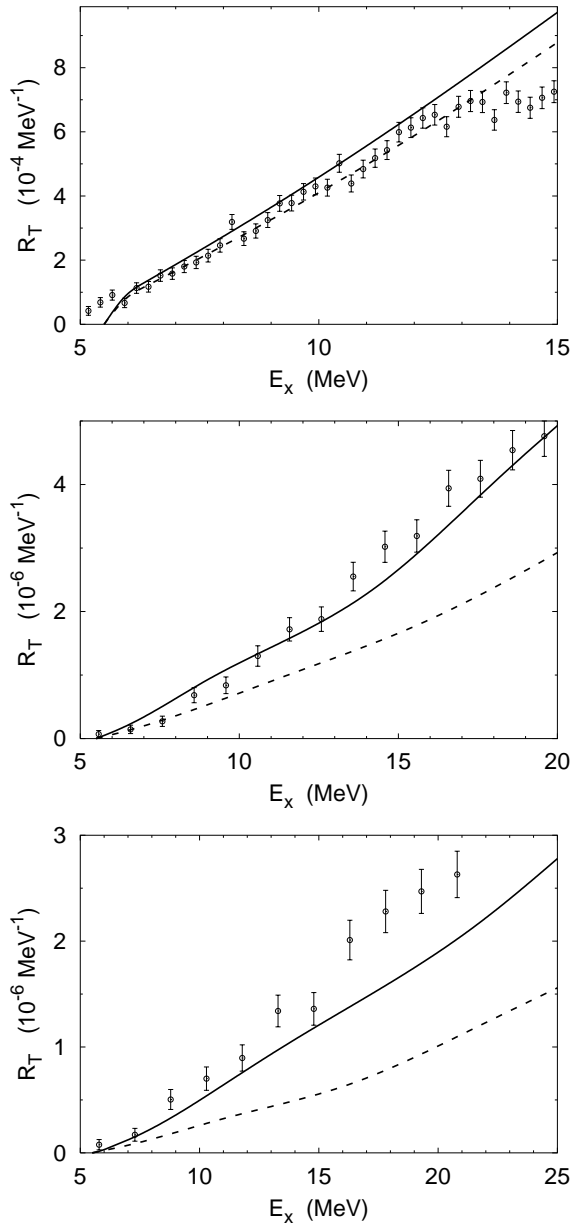}
\end{center}
\caption{\label{fig:RTt}
$\He$ inclusive transverse response $R_T$ near threshold
around the momentum transfers 
$|\mbf{Q}| = 473$, $862$ and $927\;\mathrm{MeV}/c$ from top to bottom 
as function of the excitation energy  $E_x$. 
Results of the coupled-channel potential with $\Delta$-isobar excitation 
(solid curves) are compared with reference results of the purely nucleonic 
CD-Bonn potential (dashed curves).
The experimental data are from \Ref~\cite{hicks:03a},
$|\mbf{Q}|$ being the value at threshold there.}
\end{figure}

Figures~\ref{fig:RLt} - \ref{fig:RF500} present sample results for inclusive  
longitudinal and transverse response functions 
$R_L$ and $R_T$  of unpolarized $\He$ and $\Hh$.

Figures~\ref{fig:RLt} and \ref{fig:RTt} contain results for
threshold data of sizable momentum transfer, i.e., 
$473 \leq |\mbf{Q}| \leq 927\;\mathrm{MeV}/c$;
they are given as functions of the excitation energy
$E_x= \sqrt{m_B^2 c^4 + 2m_B c^3 Q_0 - Q^2 c^2} - m_B c^2$.
The longitudinal response $R_L$ shows only a relatively small 
$\Delta$-isobar effect, not documented in the plot, but there is a clear need 
for relativistic corrections as seen already in the trinucleon charge form factors;
the same operator corrections are used there and here. 
The transverse response $R_T$ is rather well described, as the trinucleon
magnetic form factors in \Fig~\ref{fig:fm} are,
by the inclusion of the $\Delta$ isobar; the purely nucleonic calculations
of this paper as well as those of \Ref~\cite{hicks:03a}, based on a different
two-nucleon potential, fail in accounting for the experimental data at higher 
momentum transfers. 

Figures \ref{fig:RF300} and \ref{fig:RF500} contain results for the
responses at higher energy transfers including the region of the quasielastic peak.
The $\Delta$-isobar effects are rather insignificant.
The overall agreement with the experimental data is satisfactory,
though a consistent displacement of the quasielastic peak for the responses
at higher three-momentum transfer in \Fig~\ref{fig:RF500} is obvious.
The displacement occurs in all responses, but it is more discernible
 for the transverse responses whose peaks are more pronounced.
We think that that displacement is due to the use of nonrelativistic kinematics
for the baryons involved:

(1) The estimates for the position of the quasielastic peak, i.e., $Q^2/2m_N$ 
with relativistic kinematics and $\mbf{Q}^2/2m_N$ with nonrelativistic kinematics
differ just by that displacement. (2) In the plane-wave impulse approximation
of the responses by one of the present authors in \Ref~\cite{meier-hajduk:89a}
the use of relativistic kinematics in the final phase-space element is important
for the achieved agreement with experimental data.
However, a calculational improvement with respect to baryon kinematics is not 
straightforward when the full dynamics is included.

\renewcommand{\scl}{0.52}
\begin{figure*}[!]
\begin{center}
\includegraphics[scale=\scl]{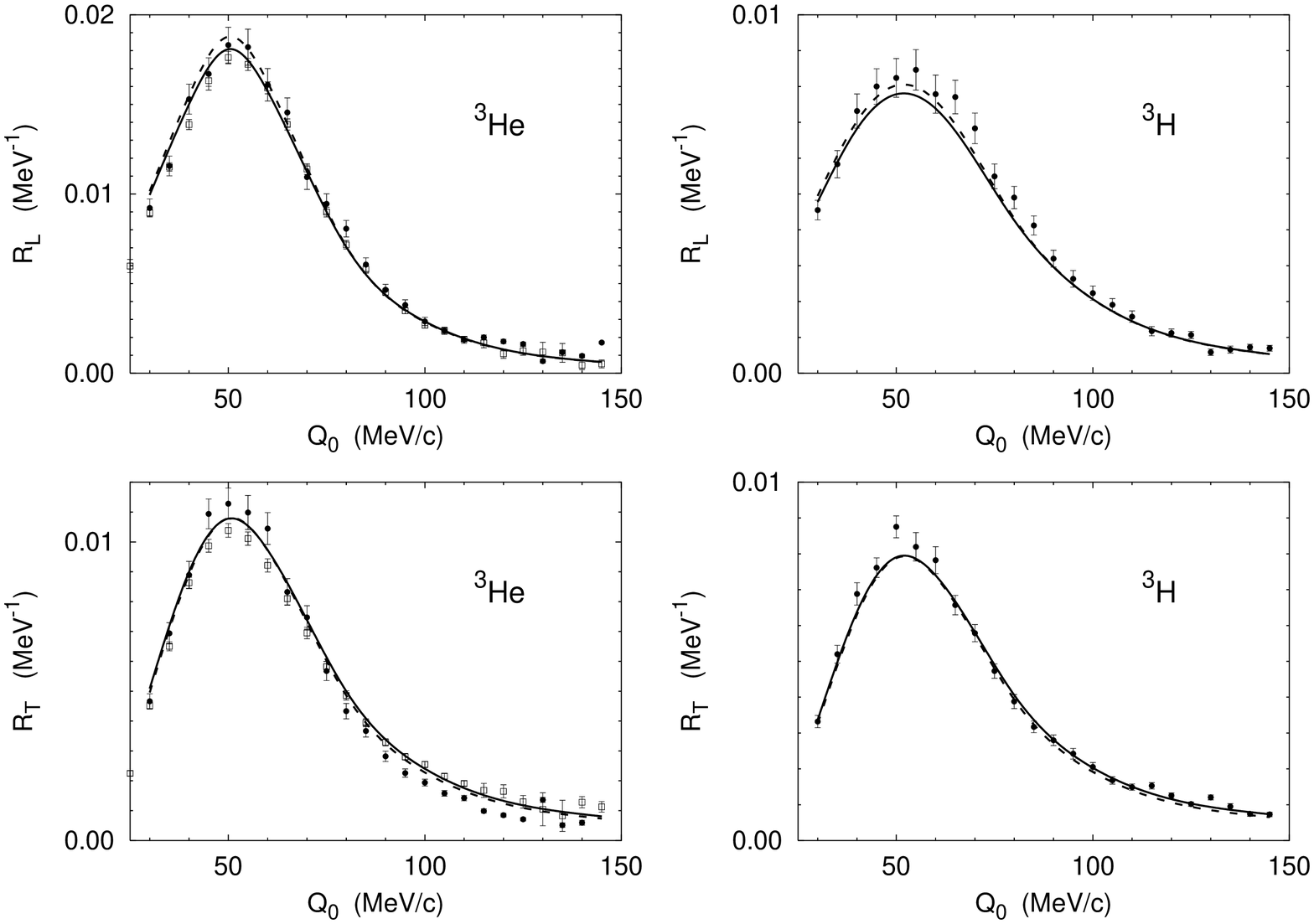}
\end{center}
\caption{\label{fig:RF300}
$\He$ and $\Hh$ inclusive longitudinal and transverse response 
functions $R_L$ and $R_T$ 
for the momentum transfer $\;|\mbf{Q}| = 300\;\mathrm{MeV}/c$ as functions
of the energy transfer $Q_0$.
Results of the coupled-channel potential with $\Delta$-isobar excitation 
(solid curves) are compared with reference results of the purely nucleonic 
CD-Bonn potential (dashed curves).
The experimental data are from \Ref~\cite{dow:88a} $(\bullet)$ 
and from \Ref~\cite{marchand:85a} $(\Box)$.}
\end{figure*}

\begin{figure*}[!]
\begin{center}
\includegraphics[scale=\scl]{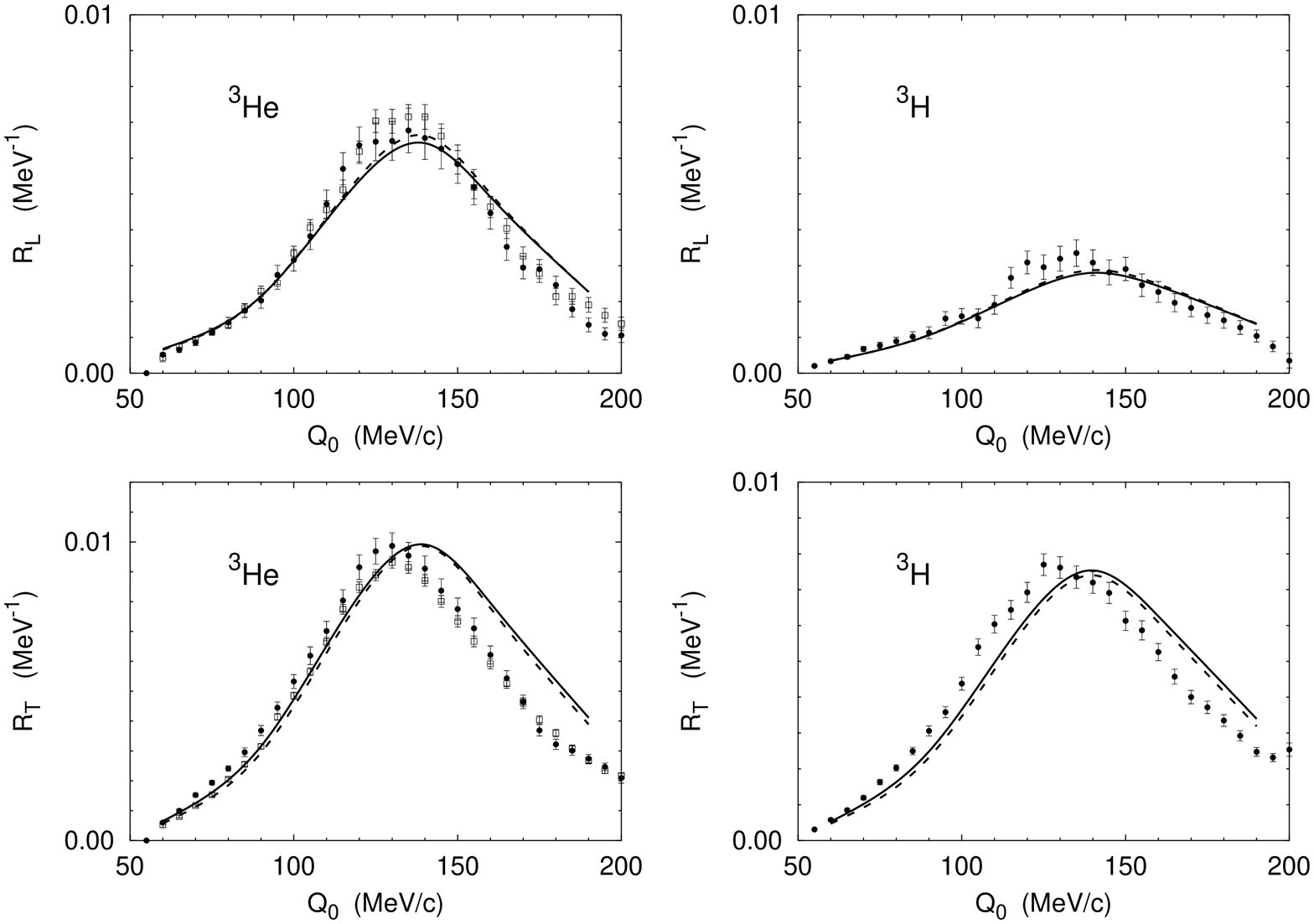}
\end{center}
\caption{\label{fig:RF500}
$\He$ and $\Hh$ inclusive longitudinal and transverse response 
functions $R_L$ and $R_T$ 
for the momentum transfer $\;|\mbf{Q}| = 500\;\mathrm{MeV}/c$ as functions
of the energy transfer $Q_0$.
Results of the coupled-channel potential with $\Delta$-isobar excitation 
(solid curves) are compared with reference results of the purely nucleonic 
CD-Bonn potential (dashed curves).
The experimental data are from \Ref~\cite{dow:88a} $(\bullet)$ 
and from \Ref~\cite{marchand:85a} $(\Box)$.}
\end{figure*}

Figures~\ref{fig:A0} and \ref{fig:A135} present results for asymmetries 
$A(\mbf{n}_B)$, measured in the experiments of \Refs~\cite{xu:00a,xiong:01a}
around the four-momentum transfer $Q^2 = 0.1$ and $0.2\,(\mathrm{GeV}/c)^2$.
The $\Delta$-isobar effects are rather insignificant. The overall agreement
between the experimental data and the theoretical predictions is rather good
for the lower four-momentum transfer, whereas at higher momentum transfer
there are some discrepancies.
\renewcommand{\scl}{0.62}
\begin{figure}[!]
\begin{center}
\includegraphics[scale=\scl]{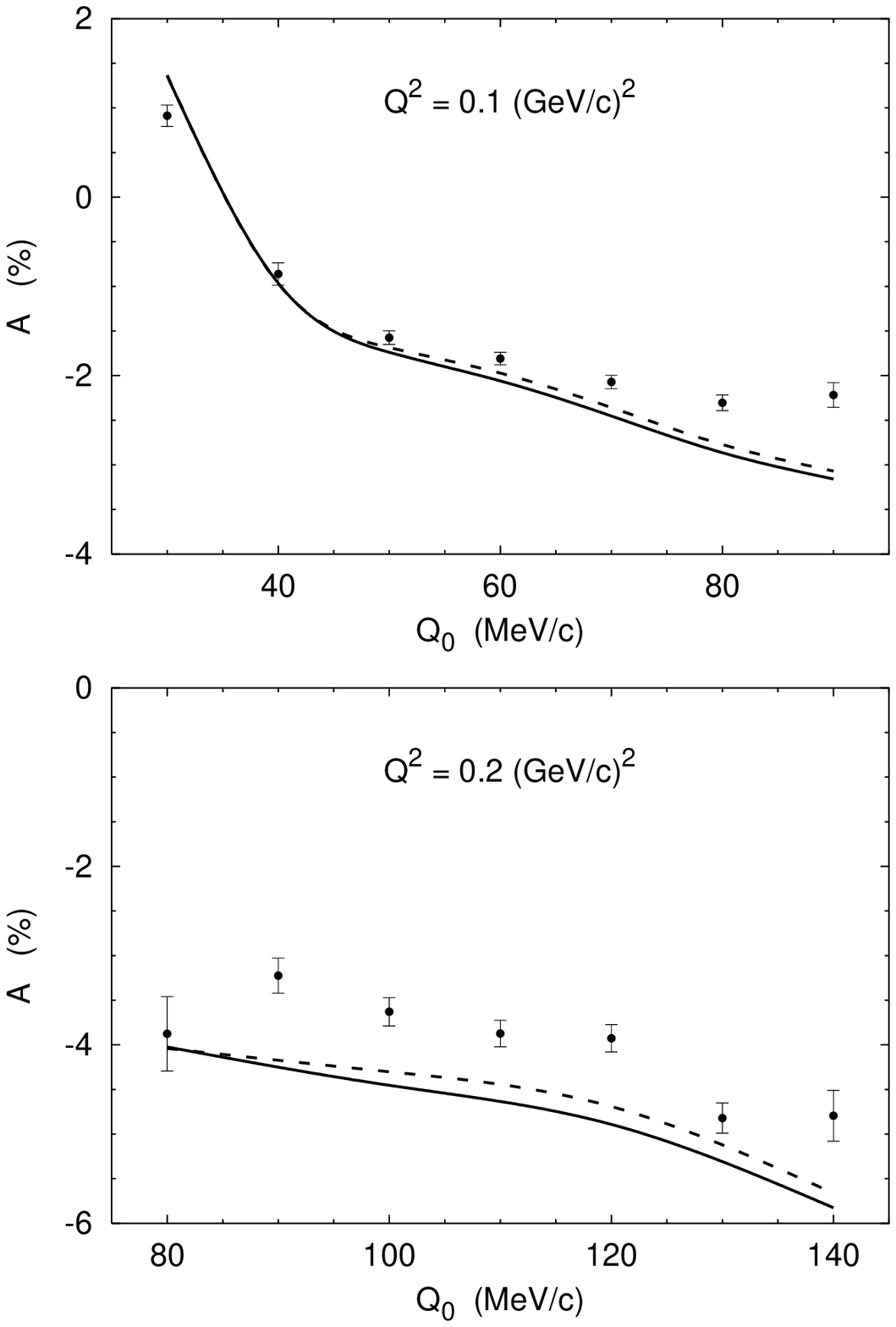} 
\end{center}
\caption{\label{fig:A0}
The inclusive asymmetry $A$ around $(\theta_B, \varphi_B) = (0^\circ, 0^\circ)$
in $\Heee$ process at four-momentum transfer $Q^2 = 0.1$ and 
$0.2\,(\mathrm{GeV}/c)^2$ as function of the energy transfer $Q_0$.
The incident electron energy is 778~MeV.
Results of the coupled-channel potential with $\Delta$-isobar excitation 
(solid curves) are compared with reference results of the purely nucleonic 
CD-Bonn potential (dashed curves).
The experimental data are from \Ref~\cite{xu:00a}. }
\end{figure}

\begin{figure}[!]
\begin{center}
\includegraphics[scale=\scl]{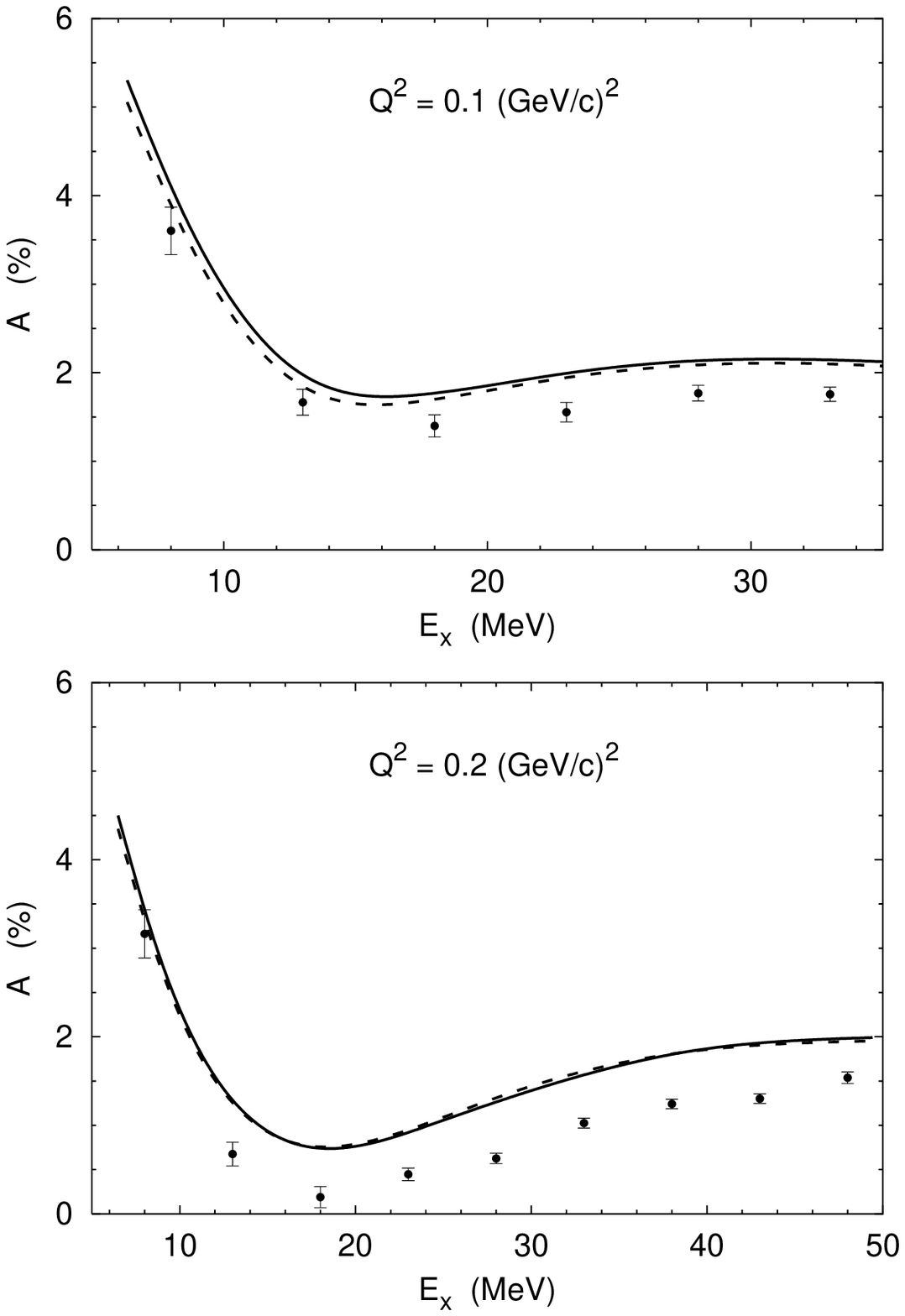}
\end{center}
\caption{\label{fig:A135}
The inclusive asymmetry $A$ around $(\theta_B, \varphi_B) = (135^\circ, 0^\circ)$
in $\Heee$ process
around the four-momentum transfer  $Q^2 = 0.1$ and $0.2\,(\mathrm{GeV}/c)^2$
as function of the excitation energy $E_x$.
The incident electron energies are 778 and 1727~MeV, and the electron scattering
scattering angles are $23.7^{\circ}$ and $15.0^{\circ}$, respectively.
Results of the coupled-channel potential with $\Delta$-isobar excitation 
(solid curves) are compared with reference results of the purely nucleonic 
CD-Bonn potential (dashed curves).
The experimental data are from \Ref~\cite{xiong:01a}.}
\end{figure}

\section{Summary and conclusions \label{sec:concl}}

The present paper completes our discussion \cite{yuan:02a,yuan:02b,deltuva:04a}
of e.m. three-nucleon processes below pion-production threshold.
Its particular focus is three-body disintegration and inclusive reactions
in inelastic electron scattering. The specialty of the description is
the use of a realistic coupled-channel potential with single
$\Delta$-isobar excitation for the initial and final hadronic states and the
use of a corresponding coupled-channel e.m. current with two-baryon
contributions. The $\Delta$-isobar effects on observables therefore result
from the effective three-nucleon force --- and a $\Delta$-modification of
the effective two-nucleon force --- and from corresponding effective
two- and three-nucleon exchange currents, all effective hadronic and e.m.
interactions mediated by the $\Delta$ isobar and based on the exchange
of all considered mesons.

We find large and beneficial $\Delta$-isobar effects for the transverse 
response in the threshold region at rather high momentum transfer 
$|\mbf{Q}| > 800\;\mathrm{MeV}/c$; all purely nucleonic
calculations fail in accounting for the corresponding experimental data. 
We also predict rather significant  $\Delta$-isobar effects
for the exclusive differential cross
section in particular kinematical regimes.
For the considered response functions and inclusive asymmetries 
up to $|\mbf{Q}| = 500\;\mathrm{MeV}/c$ 
the found $\Delta$-isobar effects are small.

We see a need for an improvement of the presented theoretical apparatus
in three respects:

(1) As already discussed in \Ref~\cite{deltuva:04a},
the employed baryonic potentials and the respective e.m. currents
are not fully consistent, the potentials being nonlocal and the currents being 
local. According to our exploratory investigation \cite{deltuva:phd} this lack 
of current conservation is practically not serious for the observables of 
electrodisintegration considered in this paper. 
Nevertheless, conceptually the development and the use of an improved and
consistent e.m. current is quite desirable.

(2) Future experiments will focus on processes above pion-production
threshold, even if only purely nucleonic channels are selected for
the explicit observation. The data of \Ref~\cite{groep:00a} corresponding to
our predictions of \Fig~\ref{fig:LQ} are only one example.
Thus, for those processes the present description of the dynamics without
an explicit pion channel is clearly insufficient; an improvement is quite
desirable. 
That improvement is also quantitatively important as \Fig~\ref{fig:LQ} proves.

(3) The four-vector e.m. current is a relativistic concept. Thus, 
the description of the hadronic initial and final states should be based
on covariant dynamic equations. Such an extension of the present 
theoretical description is highly desirable.

\begin{acknowledgments}
The authors thank H. W. Hammer and U. G. Meissner  for providing them
with their new parametrization of nucleonic form factors,
H.~Henning, E.~Jans, J.~Jourdan and I.~Sick for fruitful discussions,
and J.~Golak and I.~Nakagawa for helping them to obtain experimental data.
A.D. and L.P.Y. are partially supported by the DFG grant Sa 247/25,
J.A. by the grant GA CzR 202/03/0210 and by the projects ASCR AV0Z1048901
and K1048102. 
The numerical calculations were performed at Regionales Rechenzentrum 
f\"ur Niedersachsen.
\end{acknowledgments}

\begin{appendix}
\section{\label{app:RF} Integral equation for current matrix element}

In this appendix the current matrix elements for two- and
three-body electro disintegration of the trinucleon bound state, i.e., 
$\langle\psi^{(-)}_{\alpha} (\mbf{q}_f) \nu_{\alpha_f}| 
j^{\mu} (\mbf{Q}, \Kpl)| B \rangle $ and
$\langle\psi^{(-)}_{0} (\mbf{p}_f \mbf{q}_f) \nu_{0_f}| 
j^{\mu} (\mbf{Q}, \Kpl) | B \rangle $, are calculated.

The antisymmetrized fully correlated three-nucleon
scattering states of internal motion in nucleon-deuteron channels, i.e.,
$\langle\psi^{(-)}_{\alpha} (\mbf{q}_f) \nu_{\alpha_f}|$, and
in three-body breakup channels, i.e.,
$\langle\psi^{(-)}_{0} (\mbf{p}_f \mbf{q}_f) \nu_{0_f}|$, are
 not obtained explicitly; they are calculated only implicitly when forming 
current matrix elements.  We introduce the 
state $|X^{\mu}(Z) \rangle$, defined according to
\begin{subequations} \label{eq:X}
  \begin{align} \label{eq:Xa}
    |X^{\mu}(Z) \rangle = {} &  \big( 1+P \big) 
    j^{\mu} (\mbf{Q}, \Kpl) | B \rangle
    \nonumber \\ & 
    + P T(Z) G_0(Z) |X^{\mu}(Z) \rangle, \\  \label{eq:Xb}
    |X^{\mu} (Z)\rangle = {} & \sum_{n=0}^{\infty} [P T(Z) G_0(Z)]^n
    \nonumber \\ & \times 
    \big( 1+P \big) j^{\mu} (\mbf{Q}, \Kpl) | B \rangle,
  \end{align} 
\end{subequations}
as intermediate quantity
with $Z=E_i+i0$ being the three-particle available energy, $T(Z)$ being
the two-baryon transition matrix and $P$ the sum of the cyclic and anticyclic
permutation operators of three particles.
Equation~\eqref{eq:Xa} is an integral equation for $|X^{\mu}(Z)\rangle$, analogous
to that for the multichannel transition matrix $U(Z)$ of \Ref~\cite{deltuva:03a}:
Both equations have the same kernel, only their driving terms are different.
We therefore solve \Eq~\eqref{eq:Xa} according to the technique of 
\Ref~\cite{deltuva:03a}, summing the Neumann series~\eqref{eq:Xb} for
$|X^{\mu} (Z)\rangle$ by the Pad\'e method. 
Once $|X^{\mu} (Z)\rangle$ is calculated, 
the current matrix elements required for the description of two- and
three-body electro disintegration of the trinucleon bound state are obtained
according to 
\begin{subequations} \label{eq:X2J}
  \begin{align} \label{eq:X2Ja}
    \langle  \psi^{(-)}_{\alpha} &(\mbf{q}_f) \nu_{\alpha_f}| 
    j^{\mu} (\mbf{Q}, \Kpl)| B \rangle  
\nonumber \\  = {} & 
    { \frac{1}{\sqrt{3}} }
    \langle \phi_{\alpha} (\mbf{q}_f) \nu_{\alpha_f}| X^{\mu} (Z)\rangle, 
    \\  \nonumber
    \langle  \psi^{(-)}_{0} & (\mbf{p}_f \mbf{q}_f) \nu_{0_f}| 
    j^{\mu} (\mbf{Q}, \Kpl) | B \rangle \\ = {} & 
    { \frac{1}{\sqrt{3}} }
    \langle \phi_{0} (\mbf{p}_f \mbf{q}_f) \nu_{0_f}| 
    \big( 1+P \big) 
    \big[ j^{\mu} (\mbf{Q}, \Kpl ) | B \rangle  
    \nonumber \\ & 
    + T(Z) G_0(Z) |X^{\mu} (Z)\rangle \big].  \label{eq:X2Jb}
  \end{align} 
\end{subequations}

When calculating the inclusive response functions,
the integration over all final hadronic states is performed implicitly, 
following  the strategy of \Ref~\cite{deltuva:04a} for
calculating the total cross section of photo disintegration.
We define the general spin-dependent response function as follows, i.e.,
\begin{subequations} \label{eq:RFM}
  \begin{align} \label{eq:RFMa}
    R_{\mathcal{M}_B' \mathcal{M}_B}^{\lambda' \lambda} (Q) =  {} &
    \epsilon_{\nu}^{\ast} ({Q} \lambda') 
    \langle B \mathcal{M}_B'|[j^{\nu} (\mbf{Q}, \Kpl) ]^{\dagger} 
\nonumber \\ & \times
    \delta (E_i - H_0 - H_I) 
\nonumber \\ & \times
j^{\mu} (\mbf{Q}, \Kpl) |B \mathcal{M}_B \rangle \,
    \epsilon_{\mu} ({Q} \lambda), \\    \label{eq:RFMb}
    R_{\mathcal{M}_B' \mathcal{M}_B}^{\lambda' \lambda} (Q) = {} &
    - \frac{1}{\pi} \mathrm{Im} \big\{ \epsilon_{\nu}^{\ast} ({Q} \lambda') 
    \langle B \mathcal{M}_B'|[j^{\nu} (\mbf{Q}, \Kpl) ]^{\dagger} 
\nonumber \\ & \times
     G(\Ei) j^{\mu} (\mbf{Q}, \Kpl) |B \mathcal{M}_B \rangle \,
    \epsilon_{\mu} ({Q} \lambda) \big\} 
  \end{align} 
with the effective  polarization vectors
$\epsilon (Q \lambda=\pm 1) = \mp \frac{1}{\sqrt{2}}(0,1,\pm i,0)$ and 
$\epsilon (Q \lambda=0) = (1,0,0,0)$. The latter choice 
assumes current conservation, i.e., the longitudinal part of the 
spatial current is replaced by the charge density
$j^{0} (\mbf{Q}, \Kpl) = j^{\mu} (\mbf{Q}, \Kpl) \epsilon_{\mu} (Q \lambda=0)$,
resulting in the effective form of $\epsilon (Q \lambda=0)$ different 
from the standard one as given, e.g., in \Ref~\cite{yuan:02b}.
In contrast to the rest of this paper, we indicate the dependence on the
spin projection $\mathcal{M}_B$ of the trinucleon bound state 
in \Eqs~\eqref{eq:RFM} explicitly. Note that
all $R_{\mathcal{M}_B' \mathcal{M}_B}^{\lambda' \lambda} (Q)$ with
$\mathcal{M}_B' + \lambda' \neq \mathcal{M}_B + \lambda$ vanish.
The auxiliary state  $G(\Ei)  j^{\mu} (\mbf{Q}, \Kpl) | B \rangle $
of \Eq~\eqref{eq:RFMb} is related to  $|X^{\mu}(\Ei) \rangle$ according to 
  \begin{align}  \label{eq:RFMc}
    G(\Ei) & j^{\mu} (\mbf{Q}, \Kpl) | B \rangle
\nonumber \\ = {} &  
    \frac13 (1+P) G_0(\Ei) \big[ j^{\mu} (\mbf{Q}, \Kpl) | B \rangle  \nonumber \\ 
           & +  T(\Ei) G_0(\Ei) |X^{\mu} (\Ei)\rangle \big].
  \end{align} 
\end{subequations}

The spin-averaged longitudinal and transverse response functions $R_L(Q)$ and
$R_T(Q)$ and the spin-dependent transverse and transverse-longitudinal response
functions $R_{T'}(Q)$ and $R_{TL'}(Q)$ are calculated according to
\begin{subequations} \label{eq:RF}
  \begin{align} \label{eq:RL}
    R_L(Q) = {} &  \frac12 \mathrm{Tr} \big[ R^{00} (Q) \big] , \\ 
    R_T(Q) = {} &  \frac12  \sum_{\lambda= \pm 1}
    \mathrm{Tr} \big[ R^{\lambda \lambda} (Q) \big], \\
    R_{T'}(Q) = {} &   \frac12 \sum_{\lambda= \pm 1} \lambda
    \mathrm{Tr} \big[ R^{\lambda \lambda} (Q) \sigma_{Bz} \big], \\
    R_{TL'}(Q) = {} &   \frac12  \sum_{\lambda' \lambda} 
    \mathrm{Tr} \big[ R^{\lambda' \lambda} (Q) \sigma_{Bx} \big] .
  \end{align} 
\end{subequations}
In \Eqs~\eqref{eq:RF} traces are calculated with respect to the 
spin quantum numbers $\mathcal{M}_B$
of the trinucleon bound state; $\sigma_{Bj}$ are the ordinary 
spin-$\frac12$ particle spin operators, i.e., the Pauli matrices,
which refer in this context to the three-nucleon target. 

\end{appendix}


\begin{thebibliography}{29}
\expandafter\ifx\csname natexlab\endcsname\relax\def\natexlab#1{#1}\fi
\expandafter\ifx\csname bibnamefont\endcsname\relax
  \def\bibnamefont#1{#1}\fi
\expandafter\ifx\csname bibfnamefont\endcsname\relax
  \def\bibfnamefont#1{#1}\fi
\expandafter\ifx\csname citenamefont\endcsname\relax
  \def\citenamefont#1{#1}\fi
\expandafter\ifx\csname url\endcsname\relax
  \def\url#1{\texttt{#1}}\fi
\expandafter\ifx\csname urlprefix\endcsname\relax\def\urlprefix{URL }\fi
\providecommand{\bibinfo}[2]{#2}
\providecommand{\eprint}[2][]{\url{#2}}

\bibitem[{\citenamefont{Yuan et~al.}(2002{\natexlab{a}})\citenamefont{Yuan,
  Chmielewski, Oelsner, Sauer, and Adam~Jr.}}]{yuan:02b}
\bibinfo{author}{\bibfnamefont{L.~P.} \bibnamefont{Yuan}},
  \bibinfo{author}{\bibfnamefont{K.}~\bibnamefont{Chmielewski}},
  \bibinfo{author}{\bibfnamefont{M.}~\bibnamefont{Oelsner}},
  \bibinfo{author}{\bibfnamefont{P.~U.} \bibnamefont{Sauer}}, \bibnamefont{and}
  \bibinfo{author}{\bibfnamefont{J.}~\bibnamefont{Adam~Jr.}},
  \bibinfo{journal}{Phys.~Rev.~C} \textbf{\bibinfo{volume}{66}},
  \bibinfo{pages}{054004} (\bibinfo{year}{2002}{\natexlab{a}}).

\bibitem[{\citenamefont{Deltuva et~al.}(2004)\citenamefont{Deltuva, Yuan,
  Adam~Jr., Fonseca, and Sauer}}]{deltuva:04a}
\bibinfo{author}{\bibfnamefont{A.}~\bibnamefont{Deltuva}},
  \bibinfo{author}{\bibfnamefont{L.~P.} \bibnamefont{Yuan}},
  \bibinfo{author}{\bibfnamefont{J.}~\bibnamefont{Adam~Jr.}},
  \bibinfo{author}{\bibfnamefont{A.~C.} \bibnamefont{Fonseca}},
  \bibnamefont{and} \bibinfo{author}{\bibfnamefont{P.~U.} \bibnamefont{Sauer}},
  \bibinfo{journal}{Phys.~Rev.~C} \textbf{\bibinfo{volume}{69}},
  \bibinfo{pages}{034004} (\bibinfo{year}{2004}).

\bibitem[{\citenamefont{Machleidt}(2001)}]{machleidt:01a}
\bibinfo{author}{\bibfnamefont{R.}~\bibnamefont{Machleidt}},
  \bibinfo{journal}{Phys.~Rev.~C} \textbf{\bibinfo{volume}{63}},
  \bibinfo{pages}{024001} (\bibinfo{year}{2001}).

\bibitem[{\citenamefont{Deltuva
  et~al.}(2003{\natexlab{a}})\citenamefont{Deltuva, Machleidt, and
  Sauer}}]{deltuva:03c}
\bibinfo{author}{\bibfnamefont{A.}~\bibnamefont{Deltuva}},
  \bibinfo{author}{\bibfnamefont{R.}~\bibnamefont{Machleidt}},
  \bibnamefont{and} \bibinfo{author}{\bibfnamefont{P.~U.} \bibnamefont{Sauer}},
  \bibinfo{journal}{Phys.~Rev.~C} \textbf{\bibinfo{volume}{68}},
  \bibinfo{pages}{024005} (\bibinfo{year}{2003}{\natexlab{a}}).

\bibitem[{\citenamefont{Deltuva
  et~al.}(2003{\natexlab{b}})\citenamefont{Deltuva, Chmielewski, and
  Sauer}}]{deltuva:03a}
\bibinfo{author}{\bibfnamefont{A.}~\bibnamefont{Deltuva}},
  \bibinfo{author}{\bibfnamefont{K.}~\bibnamefont{Chmielewski}},
  \bibnamefont{and} \bibinfo{author}{\bibfnamefont{P.~U.} \bibnamefont{Sauer}},
  \bibinfo{journal}{Phys.~Rev.~C} \textbf{\bibinfo{volume}{67}},
  \bibinfo{pages}{034001} (\bibinfo{year}{2003}{\natexlab{b}}).

\bibitem[{\citenamefont{Golak et~al.}(1995{\natexlab{a}})\citenamefont{Golak,
  Kamada, Wita{\l}a, Gl\"ockle, and Ishikawa}}]{golak:95a}
\bibinfo{author}{\bibfnamefont{J.}~\bibnamefont{Golak}},
  \bibinfo{author}{\bibfnamefont{H.}~\bibnamefont{Kamada}},
  \bibinfo{author}{\bibfnamefont{H.}~\bibnamefont{Wita{\l}a}},
  \bibinfo{author}{\bibfnamefont{W.}~\bibnamefont{Gl\"ockle}},
  \bibnamefont{and} \bibinfo{author}{\bibfnamefont{S.}~\bibnamefont{Ishikawa}},
  \bibinfo{journal}{Phys.~Rev.~C} \textbf{\bibinfo{volume}{51}},
  \bibinfo{pages}{1638} (\bibinfo{year}{1995}{\natexlab{a}}).

\bibitem[{\citenamefont{Golak et~al.}(1995{\natexlab{b}})\citenamefont{Golak,
  Wita{\l}a, Kamada, H\"uber, Ishikawa, and Gl\"ockle}}]{golak:95b}
\bibinfo{author}{\bibfnamefont{J.}~\bibnamefont{Golak}},
  \bibinfo{author}{\bibfnamefont{H.}~\bibnamefont{Wita{\l}a}},
  \bibinfo{author}{\bibfnamefont{H.}~\bibnamefont{Kamada}},
  \bibinfo{author}{\bibfnamefont{D.}~\bibnamefont{H\"uber}},
  \bibinfo{author}{\bibfnamefont{S.}~\bibnamefont{Ishikawa}}, \bibnamefont{and}
  \bibinfo{author}{\bibfnamefont{W.}~\bibnamefont{Gl\"ockle}},
  \bibinfo{journal}{Phys.~Rev.~C} \textbf{\bibinfo{volume}{52}},
  \bibinfo{pages}{1216} (\bibinfo{year}{1995}{\natexlab{b}}).

\bibitem[{\citenamefont{Golak et~al.}(2002)\citenamefont{Golak, Gl\"ockle,
  Kamada, Wita{\l}a, Skibi\'nski, and Nogga}}]{golak:02a}
\bibinfo{author}{\bibfnamefont{J.}~\bibnamefont{Golak}},
  \bibinfo{author}{\bibfnamefont{W.}~\bibnamefont{Gl\"ockle}},
  \bibinfo{author}{\bibfnamefont{H.}~\bibnamefont{Kamada}},
  \bibinfo{author}{\bibfnamefont{H.}~\bibnamefont{Wita{\l}a}},
  \bibinfo{author}{\bibfnamefont{R.}~\bibnamefont{Skibi\'nski}},
  \bibnamefont{and} \bibinfo{author}{\bibfnamefont{A.}~\bibnamefont{Nogga}},
  \bibinfo{journal}{Phys.~Rev.~C} \textbf{\bibinfo{volume}{65}},
  \bibinfo{pages}{044002} (\bibinfo{year}{2002}).

\bibitem[{\citenamefont{Nemoto et~al.}(1998)\citenamefont{Nemoto, Chmielewski,
  Haidenbauer, Oryu, Sauer, and Schellingerhout}}]{nemoto:98a}
\bibinfo{author}{\bibfnamefont{S.}~\bibnamefont{Nemoto}},
  \bibinfo{author}{\bibfnamefont{K.}~\bibnamefont{Chmielewski}},
  \bibinfo{author}{\bibfnamefont{J.}~\bibnamefont{Haidenbauer}},
  \bibinfo{author}{\bibfnamefont{S.}~\bibnamefont{Oryu}},
  \bibinfo{author}{\bibfnamefont{P.~U.} \bibnamefont{Sauer}}, \bibnamefont{and}
  \bibinfo{author}{\bibfnamefont{N.~W.} \bibnamefont{Schellingerhout}},
  \bibinfo{journal}{Few-Body Systems} \textbf{\bibinfo{volume}{24}},
  \bibinfo{pages}{213} (\bibinfo{year}{1998}).

\bibitem[{\citenamefont{Hajduk et~al.}(1983)\citenamefont{Hajduk, Sauer, and
  Strueve}}]{hajduk:83a}
\bibinfo{author}{\bibfnamefont{C.}~\bibnamefont{Hajduk}},
  \bibinfo{author}{\bibfnamefont{P.~U.} \bibnamefont{Sauer}}, \bibnamefont{and}
  \bibinfo{author}{\bibfnamefont{W.}~\bibnamefont{Strueve}},
  \bibinfo{journal}{Nucl.\@ Phys.\@} \textbf{\bibinfo{volume}{A405}},
  \bibinfo{pages}{581} (\bibinfo{year}{1983}).

\bibitem[{\citenamefont{Deltuva
  et~al.}(2003{\natexlab{c}})\citenamefont{Deltuva, Chmielewski, and
  Sauer}}]{deltuva:03b}
\bibinfo{author}{\bibfnamefont{A.}~\bibnamefont{Deltuva}},
  \bibinfo{author}{\bibfnamefont{K.}~\bibnamefont{Chmielewski}},
  \bibnamefont{and} \bibinfo{author}{\bibfnamefont{P.~U.} \bibnamefont{Sauer}},
  \bibinfo{journal}{Phys.~Rev.~C} \textbf{\bibinfo{volume}{67}},
  \bibinfo{pages}{054004} (\bibinfo{year}{2003}{\natexlab{c}}).

\bibitem[{\citenamefont{Oelsner}(1999)}]{oelsner:phd}
\bibinfo{author}{\bibfnamefont{M.}~\bibnamefont{Oelsner}}, Ph.D. thesis,
  \bibinfo{school}{Universit\"at Hannover} (\bibinfo{year}{1999}),
  \urlprefix\url{http://edok01.tib.uni-hannover.de/edoks/e002/300225598.pdf}.

\bibitem[{\citenamefont{Deltuva}(2003)}]{deltuva:phd}
\bibinfo{author}{\bibfnamefont{A.}~\bibnamefont{Deltuva}}, Ph.D. thesis,
  \bibinfo{school}{Universit\"at Hannover} (\bibinfo{year}{2003}).

\bibitem[{\citenamefont{Yuan et~al.}(2002{\natexlab{b}})\citenamefont{Yuan,
  Chmielewski, Oelsner, Sauer, Fonseca, and Adam~Jr.}}]{yuan:02a}
\bibinfo{author}{\bibfnamefont{L.~P.} \bibnamefont{Yuan}},
  \bibinfo{author}{\bibfnamefont{K.}~\bibnamefont{Chmielewski}},
  \bibinfo{author}{\bibfnamefont{M.}~\bibnamefont{Oelsner}},
  \bibinfo{author}{\bibfnamefont{P.~U.} \bibnamefont{Sauer}},
  \bibinfo{author}{\bibfnamefont{A.~C.} \bibnamefont{Fonseca}},
  \bibnamefont{and} \bibinfo{author}{\bibfnamefont{J.}~\bibnamefont{Adam~Jr.}},
  \bibinfo{journal}{Few-Body Systems} \textbf{\bibinfo{volume}{32}},
  \bibinfo{pages}{83} (\bibinfo{year}{2002}{\natexlab{b}}).

\bibitem[{\citenamefont{Hammer and Meissner}(2003)}]{hammer:04a}
\bibinfo{author}{\bibfnamefont{H.~W.} \bibnamefont{Hammer}} \bibnamefont{and}
  \bibinfo{author}{\bibfnamefont{U.~G.} \bibnamefont{Meissner}}
  (\bibinfo{year}{2003}), \bibinfo{note}{hep-ph/0312081}.

\bibitem[{\citenamefont{Gari and Kr\"umpelmann}(1986)}]{gari:86a}
\bibinfo{author}{\bibfnamefont{M.}~\bibnamefont{Gari}} \bibnamefont{and}
  \bibinfo{author}{\bibfnamefont{W.}~\bibnamefont{Kr\"umpelmann}},
  \bibinfo{journal}{Phys.~Lett.~B} \textbf{\bibinfo{volume}{173}},
  \bibinfo{pages}{10} (\bibinfo{year}{1986}).

\bibitem[{\citenamefont{Kamalov et~al.}(2001)\citenamefont{Kamalov, Yang,
  Drechsel, Hanstein, and Tiator}}]{kamalov:01a}
\bibinfo{author}{\bibfnamefont{S.~S.} \bibnamefont{Kamalov}},
  \bibinfo{author}{\bibfnamefont{S.~N.} \bibnamefont{Yang}},
  \bibinfo{author}{\bibfnamefont{D.}~\bibnamefont{Drechsel}},
  \bibinfo{author}{\bibfnamefont{O.}~\bibnamefont{Hanstein}}, \bibnamefont{and}
  \bibinfo{author}{\bibfnamefont{L.}~\bibnamefont{Tiator}},
  \bibinfo{journal}{Phys.~Rev.~C} \textbf{\bibinfo{volume}{64}},
  \bibinfo{pages}{032201} (\bibinfo{year}{2001}).

\bibitem[{\citenamefont{Carlson and Schiavilla}(1998)}]{carlson:98a}
\bibinfo{author}{\bibfnamefont{J.}~\bibnamefont{Carlson}} \bibnamefont{and}
  \bibinfo{author}{\bibfnamefont{R.}~\bibnamefont{Schiavilla}},
  \bibinfo{journal}{Rev.~Mod.~Phys.} \textbf{\bibinfo{volume}{70}},
  \bibinfo{pages}{743} (\bibinfo{year}{1998}).

\bibitem[{\citenamefont{Henning et~al.}(1995)\citenamefont{Henning, Adam~Jr.\@,
  Sauer, and Stadler}}]{henning:95a}
\bibinfo{author}{\bibfnamefont{H.}~\bibnamefont{Henning}},
  \bibinfo{author}{\bibfnamefont{J.}~\bibnamefont{Adam~Jr.\@}},
  \bibinfo{author}{\bibfnamefont{P.~U.} \bibnamefont{Sauer}}, \bibnamefont{and}
  \bibinfo{author}{\bibfnamefont{A.}~\bibnamefont{Stadler}},
  \bibinfo{journal}{Phys.\@ Rev.\@ C} \textbf{\bibinfo{volume}{52}},
  \bibinfo{pages}{R471} (\bibinfo{year}{1995}).

\bibitem[{\citenamefont{Marcucci et~al.}(1998)\citenamefont{Marcucci, Riska,
  and Schiavilla}}]{marcucci:98a}
\bibinfo{author}{\bibfnamefont{L.~E.} \bibnamefont{Marcucci}},
  \bibinfo{author}{\bibfnamefont{D.~O.} \bibnamefont{Riska}}, \bibnamefont{and}
  \bibinfo{author}{\bibfnamefont{R.}~\bibnamefont{Schiavilla}},
  \bibinfo{journal}{Phys.~Rev.~C} \textbf{\bibinfo{volume}{58}},
  \bibinfo{pages}{3069} (\bibinfo{year}{1998}).

\bibitem[{\citenamefont{Sick}(2001)}]{sick:01a}
\bibinfo{author}{\bibfnamefont{I.}~\bibnamefont{Sick}},
  \bibinfo{journal}{Prog.~Part.~Nucl.~Phys.} \textbf{\bibinfo{volume}{47}},
  \bibinfo{pages}{245} (\bibinfo{year}{2001}).

\bibitem[{\citenamefont{Groep~{\it et al.}}(2001)}]{groep:00a}
\bibinfo{author}{\bibfnamefont{D.~L.} \bibnamefont{Groep~{\it et al.}}},
  \bibinfo{journal}{Phys.~Rev.~C} \textbf{\bibinfo{volume}{63}},
  \bibinfo{pages}{014005} (\bibinfo{year}{2001}).

\bibitem[{\citenamefont{Meier-Hajduk et~al.}(1989)\citenamefont{Meier-Hajduk,
  Oelfke, and Sauer}}]{meier-hajduk:89a}
\bibinfo{author}{\bibfnamefont{H.}~\bibnamefont{Meier-Hajduk}},
  \bibinfo{author}{\bibfnamefont{U.}~\bibnamefont{Oelfke}}, \bibnamefont{and}
  \bibinfo{author}{\bibfnamefont{P.~U.} \bibnamefont{Sauer}},
  \bibinfo{journal}{Nucl.\@ Phys.\@} \textbf{\bibinfo{volume}{A499}},
  \bibinfo{pages}{637} (\bibinfo{year}{1989}).

\bibitem[{\citenamefont{Retzlaff~{\it et al.}}(1994)}]{retzlaff:94a}
\bibinfo{author}{\bibfnamefont{G.~A.} \bibnamefont{Retzlaff~{\it et al.}}},
  \bibinfo{journal}{Phys.~Rev.~C} \textbf{\bibinfo{volume}{49}},
  \bibinfo{pages}{1263} (\bibinfo{year}{1994}).

\bibitem[{\citenamefont{Hicks~{\it et al.}}(2003)}]{hicks:03a}
\bibinfo{author}{\bibfnamefont{R.~S.} \bibnamefont{Hicks~{\it et al.}}},
  \bibinfo{journal}{Phys.~Rev.~C} \textbf{\bibinfo{volume}{67}},
  \bibinfo{pages}{064004} (\bibinfo{year}{2003}).

\bibitem[{\citenamefont{Dow~{\it et al.}}(1988)}]{dow:88a}
\bibinfo{author}{\bibfnamefont{K.}~\bibnamefont{Dow~{\it et al.}}},
  \bibinfo{journal}{Phys.~Rev.~Lett.} \textbf{\bibinfo{volume}{61}},
  \bibinfo{pages}{1706} (\bibinfo{year}{1988}).

\bibitem[{\citenamefont{Marchand~{\it et al.}}(1985)}]{marchand:85a}
\bibinfo{author}{\bibfnamefont{C.}~\bibnamefont{Marchand~{\it et al.}}},
  \bibinfo{journal}{Phys.~Lett.~B} \textbf{\bibinfo{volume}{153}},
  \bibinfo{pages}{29} (\bibinfo{year}{1985}).

\bibitem[{\citenamefont{Xu~{\it et al.}}(2000)}]{xu:00a}
\bibinfo{author}{\bibfnamefont{W.}~\bibnamefont{Xu~{\it et al.}}},
  \bibinfo{journal}{Phys.~Rev.~Lett.} \textbf{\bibinfo{volume}{85}},
  \bibinfo{pages}{2900} (\bibinfo{year}{2000}).

\bibitem[{\citenamefont{Xiong~{\it et al.}}(2001)}]{xiong:01a}
\bibinfo{author}{\bibfnamefont{F.}~\bibnamefont{Xiong~{\it et al.}}},
  \bibinfo{journal}{Phys.~Rev.~Lett.} \textbf{\bibinfo{volume}{87}},
  \bibinfo{pages}{242501} (\bibinfo{year}{2001}).

\end{thebibliography}

\end{document}